\documentclass[]{aa}  

\usepackage{comment}
\usepackage{graphicx}
\usepackage{txfonts}
\usepackage{siunitx}
\usepackage{multirow}
\usepackage{amsmath, amssymb}
\usepackage{subfig}
\usepackage{soul}

\newcolumntype{Q}{>{\tiny}c}
\usepackage[]{hyperref}
\hypersetup{backref=true, pagebackref=true, hyperindex=true, breaklinks=true,colorlinks=true,urlcolor=blue, linkcolor=blue, citecolor=blue,pagecolor=red, bookmarks=true, bookmarksopen=true}

\begin{document}

   \title{A correlation between accretion and outflow rates for Class II Young Stellar Objects with full and transition disks}
   %\subtitle{A multi-wavelength analysis of ALMA, ATCA, and VLA data}

   \author{ A. A. Rota
          \inst{\ref{leiden}},
N. van der Marel \inst{\ref{leiden}},
A. Garufi \inst{\ref{inaf}},
C. Carrasco-Gonz\'alez \inst{\ref{unam}},       
E. Macias \inst{\ref{eso}},
I. Pascucci \inst{\ref{tucson}},
A. Sellek \inst{\ref{leiden}},
L. Testi \inst{\ref{bologna}},
A. Isella \inst{\ref{rice}},
S. Facchini \inst{\ref{milano}}
}

    \institute{Leiden Observatory, Leiden University, P.O. Box 9513, 2300 RA Leiden, The Netherlands\label{leiden}\\ \email{rota@strw.leidenuniv.nl}
    \and
INAF - Istituto di Radioastronomia, Via Gobetti 101, I-40129, Bologna, Italy\label{inaf}
\and
Instituto de Radioastronomía y Astrofísica (IRyA-UNAM), Morelia, Michoacán 58089, Mexico\label{unam}
\and
European Southern Observatory, Karl-Schwarzschild-Strasse 2, D-85748 Garching, Germany\label{eso}
\and
Lunar and Planetary Laboratory, University of Arizona, Tucson, AZ 85721, USA
\label{tucson}
\and
Dipartimento di Fisica e Astronomia Augusto Righi, Università di Bologna, Viale Berti Pichat 6/2, Bologna, Italy\label{bologna}
\and
Department of Physics and Astronomy, Rice University, 6100 Main Street, MS-108, Houston, TX 77005, USA\label{rice}
\and
Dipartimento di Fisica, Universit\'a degli Studi di Milano, Via Celoria 16, 20133 Milano, Italy\label{milano}}
        
\titlerunning{Correlation between accretion and outflow in disks}
\authorrunning{Rota et al.}

   \date{Received -; accepted -}

% \abstract{}{}{}{}{} 
% 5 {} token are mandatory
 
\abstract{Magnetothermal (MHD) winds and jets are processes
that influence the global evolution of the star and planet formation process. They originate in a wide range of regions of protoplanetary disks ($\sim 1-30$ au) and are thought to be the primary mechanisms driving accretion onto the central star. One indirect signature of these processes is the free-free emission from ionized gas close to the star.}{We analyze a sample of 31 Class II young stellar objects (YSOs) with different types of disks:  18 full disks and 13 transition disks. All sources show evidence of excess free-free emission over the contribution of thermal dust. We investigate the origin of this emission and whether it is associated with other observables in disks with different types of substructures.}{We first analyzed a sample of objects in Taurus, exploring possible correlations with the properties of the central star, the disk, and other disk-wind tracers. We then compared our findings with a sample of transition disks, for which free-free emission was already shown to be likely associated with an MHD-wind/jet.}{We found no correlation between the detected free-free emission and either the X-ray or the [O I]6300 \r{A} line properties. We found a strong correlation between the ionized mass loss rate, as inferred from the free-free emission, and the accretion rate, suggesting that free-free emission in YSOs with full disks is associated with an MHD-wind/jet.}{The detected free-free emission in YSOs with both transition disks and full disks is likely associated with a similar mechanism, i.e. ionized gas closed to the star from an MHD-wind/jet. The free-free emission detected in transition disks shows hints of shallower correlations with accretion properties than in full disks.
Whereas the efficiency in transforming accretion into outflow might differ in transition disks and full disks, considering the correlations between free-free emission and accretion properties, this difference could simply result from a bias toward strong accretors in the transition disk sample. Therefore, additional observations of a more complete and uniform sample are necessary to determine whether this change in correlations holds only for strong accretors or for transition disks in general.
}


\maketitle

%
%-------------------------------------------------------------------

\section{Introduction}\label{intro}

A key question in the star and planet formation process is understanding the accretion process in young stellar objects (YSOs) and the evolution of protoplanetary disks that are associated with them.  Both simulations and observations have challenged the classical paradigm of viscous accretion, as the turbulence driven by magnetorotational instability (MRI) is insufficient to account for the observed accretion rates (\citealp{2023PascucciPPVII} and references therein). Instead, accretion is likely driven by radially extended magnetohydrodynamic (MHD) winds, i.e. outflowing gas from the disk atmosphere, while photevaporative (PE) winds play a crucial role at later stages of disk evolution by clearing and dispersing the disks (e.g., \citealp{2023PascucciPPVII}; \citealp{2021Lesur}). It is thus crucial for our understanding of disk evolution to study the relationship between the accretion rate onto the star and the disk wind.

From an observational point of view,  MHD disk winds exhibit a nested structure: high-velocity jets ($\sim100$ km/s), which form through the recollimation of slow ($\sim30$ km/s) inner winds, are surrounded by layers of atomic and molecular gas moving at decreasing velocities (such as recently observed with \emph{JWST}; e.g., \citealp{2024Pascucci}, \citealp{2024Nisini}).
These processes originate across a wide range of disk regions  ($\sim1-30$ au) and may significantly impact the formation of planetesimals in the disks. It is particularly interesting to compare disk-wind processes in full disks and in transition disks with large inner dust cavities \citep[$>20$ au, see e.g.][for a review]{2023vanderMarel}, as it raises the question of whether accreting material can be transported throughout the disk in a similar way as through an empty cavity  \citep{2022Martel}. Photoevaporation is unlikely to play a role in clearing such transition disks, considering their massive outer disks, their high accretion rates and large cavities \citep{2012OwenClarke,2019Picogna,2023Appelgren}, and observed disk winds in those systems are more likely to be magnetically driven.

Traditionally, collimated jets and radially extended disk winds in Class II sources are studied through high-resolution spectroscopy of optical forbidden lines, such as the [O I]6300 \r{A} line (e.g., \citealp{1995Hartigan}, \citealp{2018Fang}; \citealp{2018Nisini}). These lines are usually decomposed in a blueshifted high-velocity component (HVC) associated with a collimated fast jet, and a low-velocity component (LVC) often associated with an MHD wind (e.g., \citealp{2016Simon}, \citealp{2019Banzatti}). Based on their different line widths, the latter component can be further decomposed into a broad (BLVC) and a narrow component (NLVC).
The study of the [O I]6300 \r{A} line in a survey of young stars in Lupus, Chamaeleon, and $\sigma$ Orionis shows that the luminosity of both the LVC and the HVC (L$_\text{[O I]LVC}$ and L$_\text{[O I]HVC}$, respectively) correlates with the stellar accretion rate and the accretion luminosity, suggesting that accretion is the main driver of line luminosity. Both L$_\text{[O I]LVC}$ and L$_\text{[O I]HVC}$ show a very similar correlation with the accretion luminosity, suggesting a common mechanism for the formation of the LVC and HVC (\citealp{2018Nisini}). Moreover, a comprehensive study of 64 T Tauri Stars shows that the BLVC and NLVC kinematics correlate, suggesting that both components originate from the same MHD disk wind (\citealp{2019Banzatti}). 
Finally, in line with the weaker emission expected over time, a survey conducted in the Upper Sco region by \cite{2023Fang} shows a lower detection rate of HVCs of the [O I] emission, while the proportion of single component LVCs is increasing. 
These surveys show that a majority of transition disks do not exhibit any HVC ($\sim70-80\%$ disks without HVCs), while HVCs are not detected in only $\sim 40\%$ of full disks (\citealp{2018Fang}; \citealp{2019Banzatti}). However, the sample on which these statistics are based is biased towards low-mass M-type stars, which tend to show a lower fraction of transition disks than higher-mass stars. 
Therefore, while the combination of optical and infrared observations hints at a possible difference in the ejection mechanisms in transition and full disks, additional observations are necessary to verify this hypothesis.

Another way to study MHD-disk winds and jets is through the indirect signatures that they are expected to produce, such as free-free emission from ionized gas close to the star. Emission at centimeter wavelengths is commonly associated with winds and jets at early (Class 0/I) stages in the formation of the star (e.g., \citealp{2018Anglada}). A multi-wavelength ALMA study of transition disks revealed free-free emission close to the star, strongly suggesting the presence of outflowing gas from an ionized jet and/or from a disk wind in these evolved YSOs (\citealp{2024Rota}). This study found a strong correlation between the ionized mass-loss rate, inferred from the free-free emission, and the stellar accretion rate, suggesting that accretion onto the star in these disks is mainly driven by an MHD-wind and/or a jet. Recently, excess centimeter emission over the thermal dust contribution has also been found in a sample of Class II YSOs with full disks in the Taurus star-forming region (\citealp{2025Garufi}). This centimeter excess appears to be associated with free-free emission and correlates with the accretion rate onto the central star, again suggesting a link between the disk accretion and the outflow. However, whether the same mechanism in transition and full disks drives the accretion onto the central star is still a matter of debate (e.g., \citealp{2023ManaraPPVII}).

In this paper, we investigate the origin of the free-free emission detected in the full disk sample analyzed by \cite{2025Garufi}. Moreover, we compare the results found for full and transition disks, discussing whether accretion in these two classes of disks is driven by the same mechanism. 
In Section \ref{sample}, we describe the sample of disks analyzed in this work. In Section  \ref{results}, we analyze the origin and implications of the free-free emission detected in \cite{2025Garufi}. Section \ref{discussion} discusses our findings and compares the transition disks with full disks. Finally, we draw our conclusions in Section \ref{conclusions}.

%--------------------------------------------------------------------
\section{Sample}\label{sample}

This work analyses a sample of 31 YSOs for which multi-wavelength observations are available in the literature at millimeter and centimeter wavelengths. 
The main sample consists of 19 sources from \cite{2025Garufi} which are associated with 16 full disks (compact and sub-structured disks) and 3 transition disks with cavities larger than $20$ au. From the sample reported in \cite{2025Garufi} we exclude DQ Tau and HP Tau, for which \cite{2025Garufi} report issues in their analysis. In addition, 2 sources with full disks (TW Hya and HD 163296) are included from \cite{2021Macias} and \cite{2022Guidi}. In the comparison between full and transition disks in Section \ref{comparison} we also consider the 10 transition disks from \cite{2024Rota}.
This adds up to a sample of 18 full disks and 13 transition disks.

The spectral types of the stars included in the final sample vary from B9 to M3, the stellar masses range from $0.25$ to $2.2 \text{M}_\odot$, the stellar luminosities range from $\sim0.1$ to $\sim16 \text{L}_\odot$, and accretion rates onto the central star range from $10^{-10}$ to $10^{-6} \text{M}_\odot /\text{yr}$. Table \ref{tab:sample} reports the stellar and disk properties of all disks in the sample.
All these disks show strong evidence of non-dust emission contaminating the total detected flux in their millimeter and/or centimeter photometry, analyzed in the studies listed above, with millimeter spectral indices $\alpha < 1.5$. The detected non-dust emission ranges from 0.02 mJy to 2 mJy at 2 cm. The first two columns in Table \ref{tab:free-free} report the fluxes and spectral indices estimated from the multi-wavelength analysis reported by \cite{2024Rota}, \cite{2025Garufi}, \cite{2021Macias}, and \cite{2022Guidi}.

In addition to the sample of 31 objects analyzed in this work, we attempted the same multi-wavelength analysis conducted by \cite{2025Garufi} on a sample of southern protoplanetary disks for which 7 mm emission has been detected by ATCA \citep{2012Ubach,2017Ubach}. However, although the SED suggests in some cases a contribution from non-dust emission, it was not possible to disentangle its contribution from the dust one, due to the lack of multiple observations at long-wavelengths needed to constrain the weak non-dust emission contribution. More observations at low frequencies of these disks are needed to conduct a similar analysis as the ones reported in this work.

\begin{table*}
\begin{center}
\tiny
\caption{The sample of this work.}\label{tab:sample}
\begin{tabular}{ccccccccccc}
\hline\hline 
 Target & Disk type & d & SpT & M$_*$  & L$_*$ &  log $\text{L}_\text{acc}$&  log $\dot{\text{M}}_\text{acc}$  & log $\text{L}_\text{X}$ &inc & Ref. \\
 &  & [pc] & & [M$_\odot$] & [L$_\odot$] & [L$_\odot$] & [M$_\odot$ yr$^{-1}$] & [L$_\odot$] &[deg] & \\
\hline

BPTau & Full & 127.4 & M0.5 & 0.48 & $0.37\pm0.20$ & $-1.34\pm0.16$ & $-8.25\pm0.4$ & $-3.45\pm0.30$ &  38.2 & 1,1,1,3\\
 CIDA9 & TD & 175.1 & M1.8 & 0.66 & $0.21\pm0.20$ & $-1.59\pm0.20$ &  $-8.74\pm0.4$ & -- & 45.6 & 2,2,2 \\
 DLTau & Full & 159.9 & K5.5 & 0.90 & $0.40\pm0.20$ & $-0.35\pm0.18$ & $-7.62\pm0.4$ &  -- & 45.0 & 1,1,1\\
 DNTau & Full & 128.6 & M0.3 & 0.55 & $0.49\pm0.20$ & $-2.10\pm0.42$ & $-9.04\pm0.4$ & $-3.52\pm0.30$ & 35.2 & 1,1,1,3\\
 DOTau & Full & 138.5 & M0.3 & 0.42 & $0.37\pm0.20$ & $-0.93\pm0.25$ & $-7.80\pm0.4$ & $-4.21\pm0.30$ & 27.6 & 1,1,1,5\\

 DRTau & Full & 193.0 & K6 & 0.89 & $0.29\pm0.20$ & $-0.57\pm0.31$ & $-7.93\pm0.4$ & $-3.94\pm0.30^\ddag$ & 5.4 & 1,1,1,6\\
 DSTau & Full & 158.4 & M0.4 & 0.61 & $0.29\pm0.20$ &  $-1.46\pm0.18$ & $-8.57\pm0.4$ & -- & 65.2 & 1,1,1\\
 FTTau & Full & 130.2 & M2.8 & 0.25 & $0.05\pm0.20$ &  $-1.93\pm0.21$ &$-8.92\pm0.4$ & $-4.64\pm0.30^\ddag$ & 35.5 & 1,1,1,6\\
 GITau & TD & 129.4 & M0.4 & 0.55 & $0.27\pm0.20$ & $-1.37\pm0.20$ & $-8.43\pm0.4$ & $-3.82\pm0.30$ & 43.8 & 1,1,1,5\\
 GKTau & Full & 129.1 & K7.5 & 1.20 & $0.87\pm0.20$ & $-1.39\pm0.17$ & $-8.67\pm0.4$ &  $-3.42\pm0.30$ & 40.2 & 1,1,1,3\\
 GOTau & Full & 142.4 & M2.3 & 0.42 & $0.20\pm0.20$ & $-2.00\pm0.20$ & $-8.92\pm0.4$ &  $-4.19\pm0.30$ & 53.9 & 3,3,3,3\\
 HOTau & Full & 164.5 & M3.2 & 0.5 & $0.16\pm0.20$ & $-2.00\pm0.20$ & $-9.13\pm0.4$ & -- & 55.0 & 4,4,4\\
 
 HQTau & Full & 161.4 & K2.0 & 2.00 & $5.83\pm0.20$ & $-1.24\pm0.17$ & $-8.43\pm0.4$ & $-2.86\pm0.30$ & 53.8 & 1,1,1,3\\
 IPTau & TD & 129.4 & M0.6 & 0.51 & $0.31\pm0.20$ & $-2.16\pm0.28$ & $-9.14\pm0.4$ & -- & 45.2 & 1,1,1\\
 IQTau & Full & 131.5 & M1.1 & 0.55 & $0.19\pm0.20$ & $-1.40\pm0.26$ & $-8.54\pm0.4$ & $-3.94\pm0.30^\ddag$ & 62.1 & 1,1,1,6\\
 MWC480 & Full & 156.2 & A4.5 & 1.71 & $10.79\pm0.20$ & $-0.19\pm0.27$ & $-7.64\pm0.4$ & $-3.82\pm0.30^\ddag$ & 36.5 & 1,1,1,6\\
 UZTauE & Full & 1231.1 & M1.9 & 0.41 & $0.45\pm0.20$ & $-1.27\pm0.25$ & $-8.05\pm0.4$ & $-4.40\pm0.30^\ddag$ & 56.1 & 1,1,1,6\\
 V409Tau & Full & 129.7 & M0.6 & 0.43 & $0.24\pm0.20$ & $-2.64\pm0.19$ & $-9.58\pm0.4$ & -- & 69.3 & 1,1,1\\
 V836Tau & Full & 167.0 & M0.8 & 0.28 & $0.63\pm0.20$ & $-1.93\pm0.20$ & $-8.37\pm0.4$ & $-3.19\pm0.30$ & 43.1 & 1,1,1,5\\
 TW Hya & Full & 59.5 & M0.0 & 0.6 & $0.3\pm0.20$ & $-1.35\pm0.25$ & $-8.65\pm0.22$ & -- & 5. & 9,11,11 \\

HD163296 & Full & 101.2 & A0 & 1.9 & $16.0\pm0.20$ & $0.73\pm0.16$ & $-6.79\pm0.15$ & -- & 5. & 12,8,8 \\

  AB Aur  & TD     & 163     & A0   & 2.2   &  $1.61^{+0.19}_{-0.21}$ & $1.32^{+0.14}_{-0.15}$  & $-6.13^{+0.25}_{-0.27}$ & $-3.70\pm0.30^\ddag$ & 23 &  7,8,8,6  \\
 GGTau AA/Ab  & TD    & 149     &  K7  &  0.7  & $0.20\pm0.20$ & $0.024\pm0.25\dagger$   & $-7.3\pm0.5$ & $-3.94\pm0.30^\ddag$ & 36 & 9,..$\dagger$,9,6   \\
 HD100453  & TD &   104    & A9   & 1.5   &  $0.85^{+0.08}_{-0.09}$ & $-0.35^{+0.28}_{-0.32}$  & $-7.79^{+0.29}_{-0.33}$ & -- & $30$ & 8,8,8  \\
 HD100546 & TD & 110     & B9   & 2.2   &  $1.45^{+0.11}_{-0.11}$ & $0.97\pm0.14$ & $-6.59^{+0.15}_{-0.12}$ & $-4.36\pm 0.30^\ddag$ & 42 & 8,8,8,6  \\
 HD135344B & TD & 136 & F5 & 1.6 & $0.94^{+0.13}_{-0.14}$ & $0.09^{+0.21}_{-0.22}$& $-7.25^{+0.21}_{-0.23}$ & -- & 12 & 8,8,8,6 \\
 HD142527 & TD   &  157    & F6   &  2.3  &  $1.39^{+0.09}_{-0.10}$ & $0.65^{+0.18}_{-0.19}$  & $-6.61^{+0.15}_{+0.18}$ & $-3.98\pm0.30^\ddag$ & 27 & 8,8,8,6  \\
 HD169142 & TD    & 114     & A5   & 2   & $1.31^{+0.12}_{-0.22}$ & $0.59\pm0.15$  & $-7.09^{+0.26}_{-0.29}$ & $-4.68\pm0.30^\ddag$ & $12$ & 7,8,8,6  \\
 MWC758 & TD    & 160     &  A7  &  1.6  &  $1.04^{+0.12}_{-0.08}$ & $0.41^{+0.17}_{-0.18}$  &  $-7.00^{+0.24}_{-0.22}$ & $-4.40\pm0.09$ & 21 & 7,8,8,10 \\
 SR24S & TD & 114     & K1   & 1.5   &  $0.40\pm0.20$ & $0.10\pm0.25\dagger$ &  $-7.15\pm0.5$ & -- & $46$ & 9,..$\dagger$,9  \\
 TCha & TD &  110     & K0   & 1.2   & $0.11\pm0.20$ & $0.012\pm0.25\dagger$  & $-8.4\pm0.5$ & $-3.11\pm0.30^\ddag$ & 73 & 9,..$\dagger$,9,6  \\
   \\
\hline
\end{tabular}
\end{center}
\tablefoot{Columns show the disk type, the distance d, the spectral type SpT, the stellar mass M$_*$ and luminosity L$_*$, the accretion luminosity $\text{L}_\text{acc}$ and the accretion rate onto the central star $\dot{\text{M}}_\text{acc}$, the X-ray luminosity, and the outer disk inclination inc. The last column reports the references for the stellar luminosity, the accretion luminosity, the accretion rates, and the X-ray luminosity. Distances are from Gaia DR2 (\citealp{2018gaiaCollabDR2}). The inclinations of the outer disks are from \citealp{2020FrancisAndVDMarel} and \cite{2019Long}. The uncertainty on the stellar luminosity is assumed to be 0.20 dex, based on the typical uncertainty estimated by \cite{2020Wichittanakom}. $\dagger$ The accretion luminosity on these targets is calculated starting from the reported accretion rate and using equation (2) in \cite{2020Wichittanakom}, with uncertainties assumed to be 0.25 dex (\citealp{2023ManaraPPVII}). $\ddag$ X-ray luminosities derived correcting the fluxes reported in Table 4 in \citealp{2019Dionatos} for the distance of each target and assuming a 0.30 dex uncertainty (\citealp{2007Gudel}).
References: 1) \cite{2022Gangi}, 2) \cite{2024Harsono}, 3) \cite{2016Simon}, 4) \cite{2010Ricci}, 5) \cite{2020Pascucci} 6) \citealp{2019Dionatos}, 7) \citealp{2018AVioque}, 8) \citealp{2020Wichittanakom}, 9) \citealp{2020FrancisAndVDMarel}, 10) \citealp{2023Ryspaeva}, 11) \citealp{2023Herczeg}, 12) \citealp{2021Varga}.}

\end{table*}

\begin{table*}
\begin{center}
\small
\caption{Summary of the free-free fluxes at 2 cm, the spectral indices of the free-free emission, and the ionized mass loss rate of the outflow as inferred from the free-free emission.}\label{tab:free-free}
\begin{tabular}{ccccc}
\hline\hline 
 Target &  Free-free flux at 2 cm [mJy] &  $\alpha_\text{ff}$ & $\dot M_\text{i} [M_ \odot /\text{yr}]$ & Ref\\ \hline

BPTau    & $0.07\pm0.02$  & $0.34^{+1.06}_{-0.70}$ & $(2.80\pm 6.84)10^{-10}$ & 1\\
DLTau    & $0.10\pm0.05$  & $0.80^{+0.80}_{-1.27}$ & $(2.05\pm5.26)10^{-9}$ & 1\\
DNTau    & $0.07\pm0.05$  & $0.85^{+0.87}_{-1.25}$ & $(1.02\pm2.80)10^{-9}$ & 1\\
DOTau    & $0.16\pm0.05$  & $0.95^{+0.82}_{-1.12}$ & $(2.65\pm7.48)10^{-9}$& 1\\
DRTau    & $0.10\pm0.04$  & $1.13^{+0.65}_{-1.43}$ & $(1.27\pm6.60)10^{-8}$ & 1\\
DSTau    & $0.03\pm0.01$  & $0.87^{+0.76}_{-1.30}$ & $(0.68\pm1.84)10^{-9}$ & 1\\
FTTau    & $0.07\pm0.04$  & $1.04^{+0.73}_{-1.31}$ & $(1.30\pm4.67)10^{-9}$ & 1\\
GKTau    & $0.05\pm0.02$  & $0.72^{+0.64}_{-1.06}$ & $(0.86\pm1.73)10^{-9}$ & 1\\
GOTau    & $0.05\pm0.03$  & $0.38^{+1.20}_{-0.83}$ & $(2.35\pm6.41)10^{-10}$ & 1\\
HOTau    & $0.05\pm0.02$  & $0.71^{+0.97}_{-1.10}$ & $(0.71\pm1.76)10^{-9}$ & 1\\
HQTau    & $0.06\pm0.02$  & $0.73^{+0.59}_{-1.13}$ & $(1.74\pm3.62)10^{-9}$ & 1\\
IQTau    & $0.06\pm0.03$  & $0.57^{+1.12}_{-0.94}$ & $(0.47\pm1.15)10^{-9}$ & 1\\
MWC480   & $0.24\pm0.08$  & $0.91^{+0.78}_{-1.23}$ & $(0.73\pm2.02)10^{-8}$ & 1\\
UZTauE   & $0.14\pm0.08$  & $0.81^{+0.68}_{-0.45}$ & $(1.17\pm1.68)10^{-9}$ & 1\\
V409Tau  & $0.02\pm0.01$  & $0.84^{+0.85}_{-1.26}$ & $(3.48\pm9.35)10^{-10}$ & 1\\
V836Tau  & $0.12\pm0.04$  & $0.64^{+1.05}_{-0.99}$ & $(0.96\pm2.30)10^{-9}$ & 1\\
TWHya    & $0.28\pm0.62$  & $0.56^{+0.40}_{-0.40}$ & $(0.79\pm1.51)10^{-9}$ & 2\\
HD163296 & $1.69\pm0.40$  & $0.21^{+0.07}_{-0.07}$ & $(2.68\pm0.81)10^{-9}$ & 3\\

CIDA9    & $0.02\pm0.01$  & $0.58^{-1.08}_{+0.98}$ & $(3.75\pm9.18)10^{-10}$ & 1\\
GITau   & $0.04\pm0.02$  & $1.04^{-1.31}_{+0.62}$ & $(1.22\pm4.10)10^{-9}$ & 1\\
IPTau   & $0.03\pm0.02$  & $0.56^{-1.04}_{+0.92}$ & $(2.84\pm6.72)10^{-10}$ & 1\\

 GG Tau & $0.22\pm0.32$ & $[-0.35,0.69](0.17)^{*}$ & $(0.57\pm1.29)10^{-9}$ & 4\\
 HD100546 & $0.48\pm0.37$ & $[0.07,0.57](0.32)^{*}$ & $(1.85\pm1.69)10^{-9}$ & 4\\
 HD142527 & $0.0067\pm0.0088$ & $[0.66,1.63](1.14)^{*}$ & $(1.41\pm3.82)10^{-9}$ & 4\\
 AB Aur & $0.21\pm0.16$ & $<0.80\pm0.16$ & $(6.59\pm4.60)10^{-9}$ & 4\\
 HD100453 & $0.018\pm0.008$ & $<1.43\pm0.08$ & --$^\dagger$ & 4\\

 HD169142 & $0.03\pm0.02$ & $<0.62\pm0.11$ & $(6.55\pm3.20)10^{-10}$ & 4\\
 HD135344B & $0.005\pm0.019$ & $<0.96\pm0.71$ & $(1.03\pm4.65)10^{-9}$ & 4\\
 MWC758 & $0.053\pm0.097$  & $<0.47\pm0.33$ &  $(0.92\pm1.46)10^{-9}$ & 4\\ 
 SR24S & $0.13\pm0.03$ & $<0.59\pm0.05$ & $1.17\pm0.24)10^{-9}$ & 4\\
TCha & $0.34\pm0.18$ & $<0.14\pm0.12$ & $(5.30\pm3.39)10^{-10}$ & 4\\

\hline

\end{tabular}
\end{center}
\tablefoot{References for free-free fluxes and spectral indices: 1) \cite{2025Garufi}, 2) \cite{2021Macias}, 3) \cite{2022Guidi}, 4) \cite{2024Rota}. $^*$We report the confidence interval for the spectral index of the central emission in transition disks, while the average value in the interval that we used to calculate the ionized mass loss rate is reported in brackets. See \cite{2024Rota} for further details. $^\dagger$ The ionized mass loss rate cannot be computed for this target since the spectral index is $>1.3$ and the denominator in the first line in Equation \ref{eq:Mion} is negative (see \citealp{2024Rota} for further details).}

\end{table*}

%--------------------------------------------------------------------

\section{Results}\label{results}

We start with the analysis of the non-dust emission detected at radio wavelengths in a sample of protoplanetary disks in Taurus, reported in \cite{2025Garufi}, in order to constrain its origin. This sample consists of 19 disks, 8 of which do not show any evidence of sub-structures and are less than $\sim50$ au in radius and thus classified as compact disks (e.g., \citealp{2019Long}).  The remaining 11 disks show sub-structures in their dust distribution and 3 out of these 11 disks are transition disks. Also, TW Hya and HD163296 are added to the full disk sample with properties from the literature (\citealp{2021Macias}; \citealp{2022Guidi}).

\begin{figure} 
   \centering
   \includegraphics[width=0.4\textwidth]{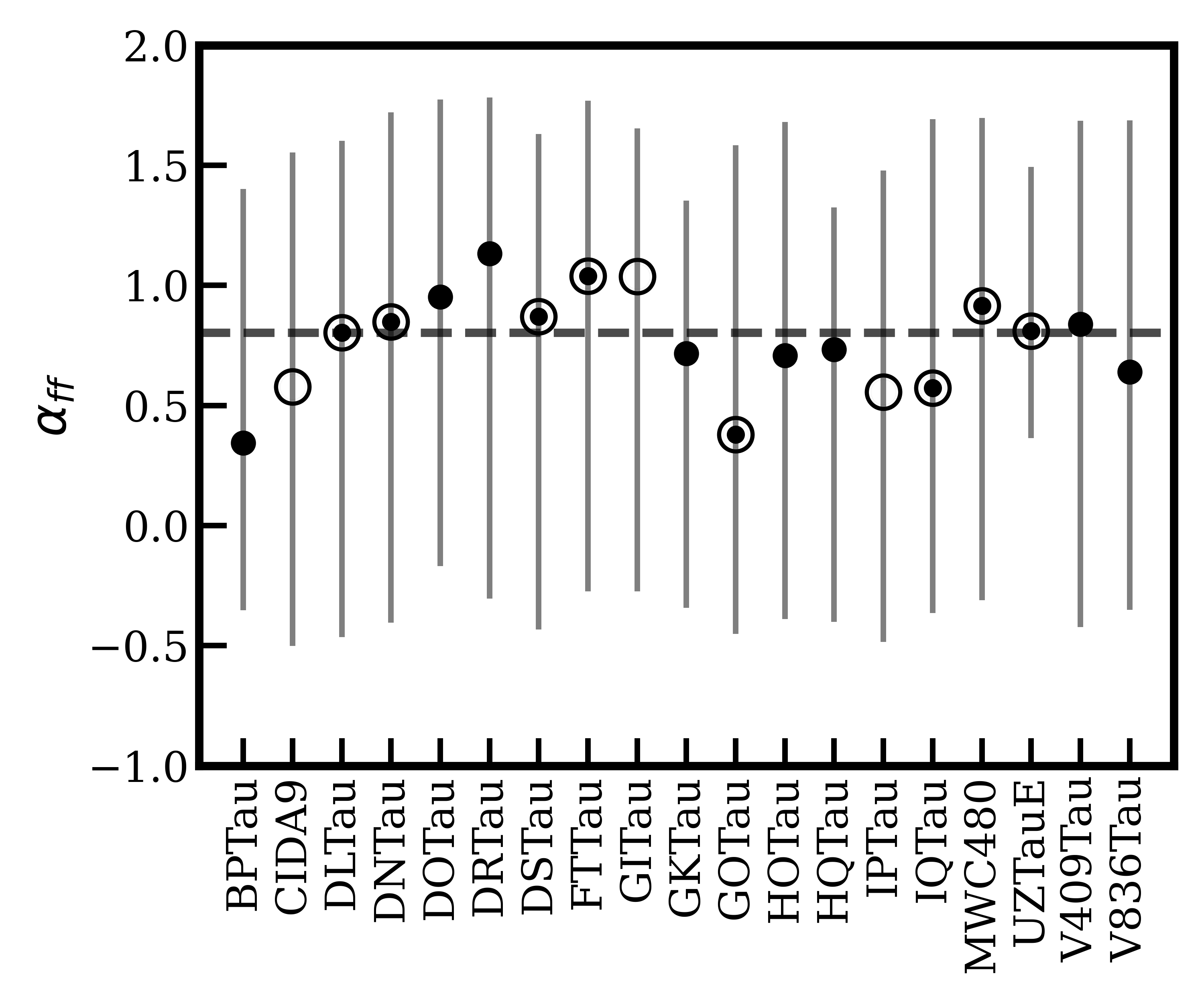}
      \caption{Spectral index of the free-free emission estimated by \cite{2025Garufi} for compact disks ($\cdot$), sub-structured disks ($\odot$), and transition disks ($\circ$). The dashed line shows the median in the sample.}
    \label{fig:alpha_freefree}  
\end{figure}

\cite{2025Garufi} present a multi-wavelength analysis of these disks, conducted including observations from VLA (Q, K, Ka, Ku, and C bands), ALMA (at 1.33 mm), IRAM Plateau de Bure Interferometer (at 2.7 mm), CSO (at 1.6 mm), and SCUBA (at 850 $\mu$m). On average, their analysis constrains a non-dust emission contribution of 65\% of the total flux at 2 cm, with the spectral indices $\alpha_\text{ff}$ spanning from 0.3 to 1.1, with a mean of 0.76 and median 0.80 (Figure \ref{fig:alpha_freefree}). These flat spectral indices are not compatible with thermal dust emission and are commonly associated with thermal free-free emission from ionized gas close to the star (e.g., \citealp{2018Anglada}). We note that other non-dust emission mechanisms, such as emission from high-velocity electrons interacting with the magnetic field, may contribute to a fraction of the detected flux (e.g., \citealp{1982Dulk}, \citealp{2018Anglada}). However, since this fraction cannot be determined and can vary from target to target and since the analysis conducted by \cite{2025Garufi} strongly suggests that free-free emission is the dominant mechanism at cm wavelengths in these objects, hereafter we refer to this emission as free-free emission.

\begin{figure} 
 \centering
\includegraphics[width=0.4\textwidth]{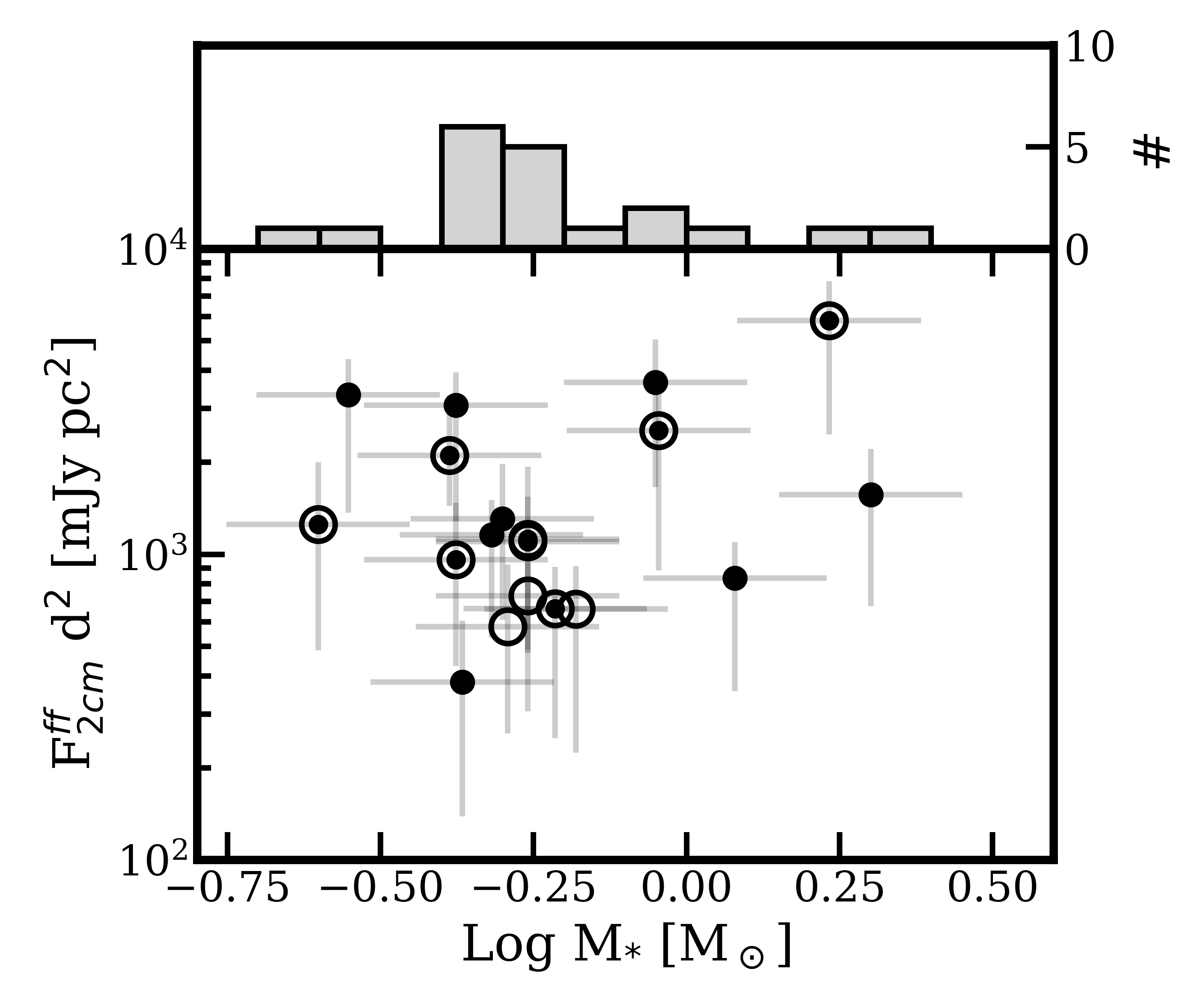}
   \centering
   \includegraphics[width=0.4\textwidth]{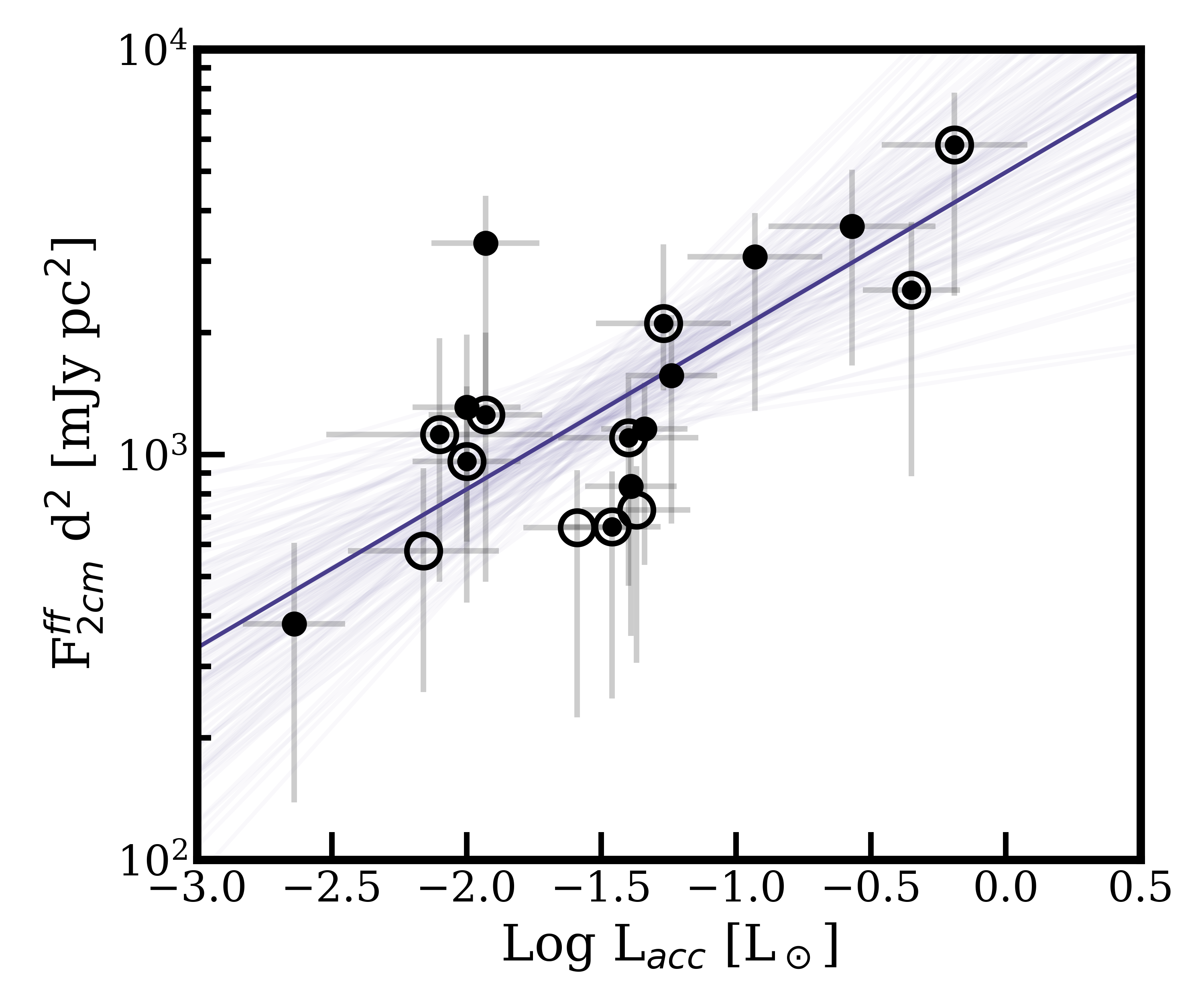}
      \caption{ \textit{Top:} Stellar mass as a function of the free-free luminosity at 2 cm estimated by \cite{2025Garufi} and histogram counts of the stellar mass. \textit{Top:} Accretion luminosity as a function of the free-free luminosity at 2 cm estimated by \cite{2025Garufi}. In both panels, compact disks ($\cdot$), sub-structured disks ($\odot$), and transition disks ($\circ$) are shown.}
    \label{fig:freefreeVSaccLumANDmass}  
\end{figure}

The free-free emission at 2 cm, $F_{\text{2cm}}^\text{ff}$, scaled at 140 pc, strongly correlates with the stellar accretion rates determined from UV excess, with a Pearson coefficient of $r=0.66$ (see Figure 2 in \citealp{2025Garufi}). 
No clear correlation between the estimated free-free luminosity at 2 cm and the stellar mass is observed ($r=0.24\pm0.24$, top panel in Figure \ref{fig:freefreeVSaccLumANDmass}) although we note that the range of stellar masses in our sample is relatively narrow. On the other hand, we found a correlation between the stellar accretion rate and the stellar mass (($r=0.77\pm0.12$), in agreement with the well-established relation reported by spectroscopic surveys (e.g., \citealp{2014Alcala}; \citealp{2016Hartmann}).
Finally, we found a clear correlation between the free-free luminosity at 2 cm and the accretion luminosity,
with $r=0.72 \pm 0.17$ (p-value $p= 5 \times 10^{-4}$, bottom panel in Figure \ref{fig:freefreeVSaccLumANDmass}), and with the following relationship\footnote{The linear regression was performed with  \texttt{linmix} (\citealp{2007Kelly})}:

\begin{equation}
    \Bigg( \frac{F_{\text{2cm}}^\text{ff}}{\text{mJy pc}^2} \Bigg) = 10^{(3.70\pm0.19)} \Bigg( \frac{L_\text{acc}}{\text{L}_\odot} \Bigg)^{(0.39\pm0.13)}.
\end{equation}

\subsection{Photoevaporative Wind Or Jet/MHD-Wind?}

The detected free-free emission can be associated with gas from a disk wind (e.g., \citealp{2012Pascucci}), an ionized jet (e.g., \citealp{2018Anglada}), or from both (e.g., \citealp{2016Macias}). In this section, we explore the possibility that the detected free-free emission is associated with a photoevaporative disk wind or a MHD-wind/jet.

\subsubsection{Accretion luminosity and X-ray luminosity}

If the free-free emission is associated with ionization by stellar EUV and/or X-ray photons, the free-free luminosity is expected to be linearly correlated with the stellar EUV photon luminosity ($\Phi_\text{EUV}$, fully-ionized wind case) and/or with the locally incident X-ray photon luminosity ($L_\text{X}$, partially ionized case; \citealp{2012Pascucci}). In this analysis, we extrapolate the observed free-free emission from 2 to 3.5 cm using the derived spectral indices in order to compare with model estimates.

\begin{figure} 
   \centering
   \includegraphics[width=0.4\textwidth]{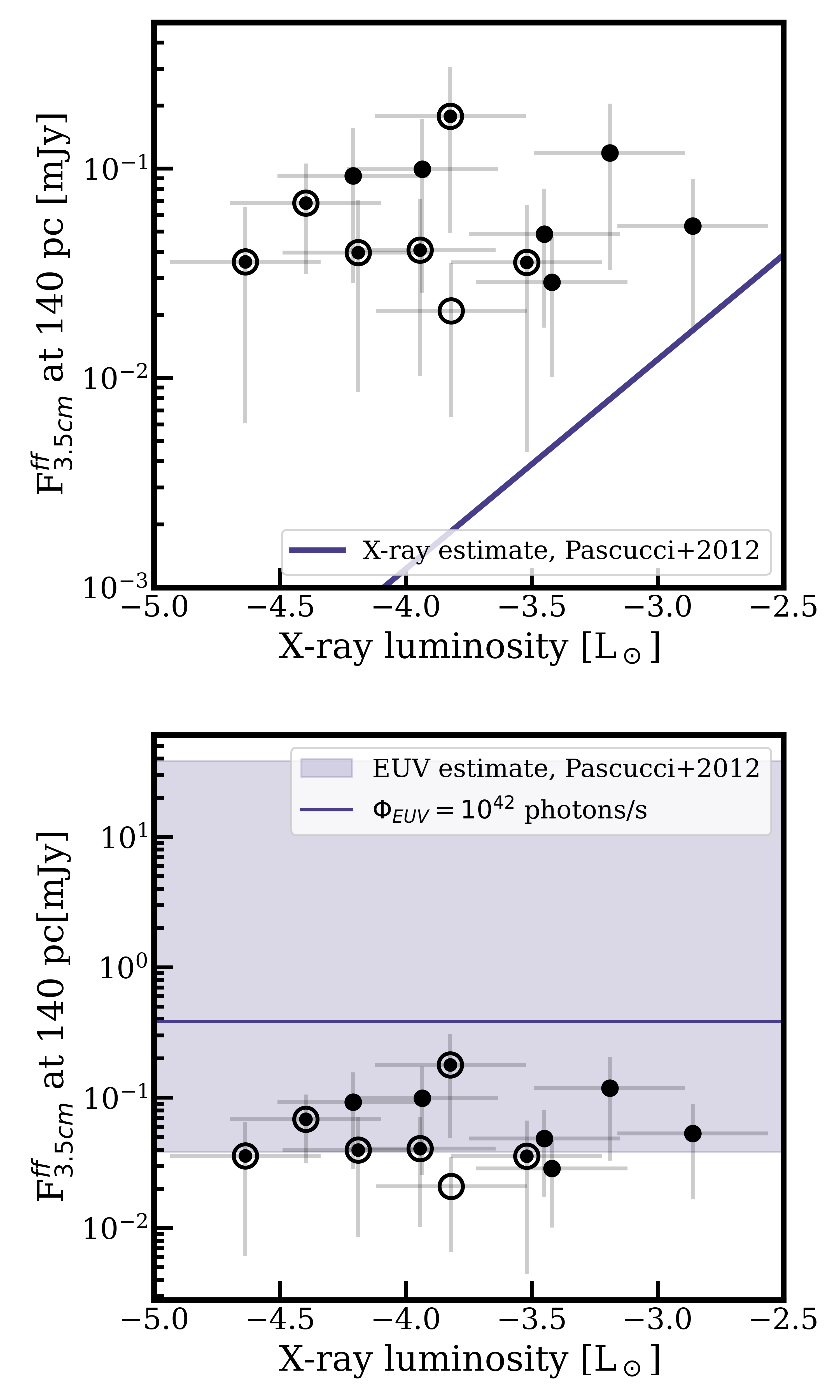}
      \caption{X-ray luminosity as a function of the free-free luminosity at 3.5 cm for compact disks ($\cdot$), sub-structured disks ($\odot$), and transition disks ($\circ$). \textit{Top:} The purple line shows the expected free-free luminosity associated with ionization by X-ray photons estimated following Equation (3) in \cite{2012Pascucci}.  \textit{Bottom:} The colored region shows the expected free-free luminosity associated with ionization by EUV photons as estimated by \cite{2012Pascucci} assuming stellar EUV luminosity $\Phi_\text{EUV} = 10^{41} - 10^{44}$ photons/s. The case where  $\Phi_\text{EUV} = 10^{42}$ photons/s is shown with a purple line. }
    \label{fig:freefreeVSxray_pascucci}  
\end{figure}

As shown by the top panel in Figure \ref{fig:freefreeVSxray_pascucci}, no correlation between the X-ray luminosity and the free-free emission extrapolated to 3.5 cm and scaled to 140 pc is found ($r=-0.3\pm0.3$).
Moreover, the top panel in the figure shows with a purple line the expected 3.5 cm free-free emission in case the emission is associated with ionization by X-ray photons (see Equation (3) in \citealp{2012Pascucci}). In most cases, the detected free-free emission extrapolated to 3.5 cm is not compatible with the expectations from \cite{2012Pascucci}. The absence of correlation and the observed excess over the expected 3.5 cm flux suggest that the observed free-free emission is unlikely to be associated with an X-ray driven photoevaporative wind. 

 Since the EUV photons are easily absorbed by the hydrogen atoms in the interstellar medium, direct measurements of the stellar EUV luminosity $\Phi_\text{EUV}$ are difficult. Therefore, a direct comparison between the detected free-free emission and the $\Phi_\text{EUV}$ is not possible. However, \cite{2012Pascucci} estimate the free-free flux at 3.5 cm from a fully ionized layer to range between 38.5 $\mu$Jy and 38.5 mJy at 140 pc, in case of $\Phi_\text{EUV} = 10^{41}$ photons/s or $\Phi_\text{EUV}=10^{44}$ photons/s, respectively. As shown from the colored region in the bottom panel in Figure \ref{fig:freefreeVSxray_pascucci}, the observed fluxes at 3.5 cm appear, in general, consistent with free-free emission associated with a fully-ionized disk wind. This is already suggested by the observed correlation between the accretion luminosity and the free-free luminosity (bottom panel in Figure \ref{fig:freefreeVSaccLumANDmass}), since the $\Phi_\text{EUV}$ is in turn expected to correlate with the accretion luminosity (\citealp{2016ErcolanoOwen}). However, photoevaporative disk winds are expected to be characterized by optically thin emission with a negative spectral index of $-0.1$ (\citealt{2012Pascucci}). On the other hand, flat or positive spectral indices (as the ones estimated for this sample) are expected to be associated with partially optically thick free-free emission due to other mechanisms, such as collimated ionized outflows and jets (\citealp{2018Anglada}).

\subsubsection{Optical Forbidden Lines}

Optical forbidden lines, such as the [O I]6300 $\r{A}$, have been used in different studies as diagnostics for MHD-winds and jets (e.g., \citealp{2024Nisini}; \citealp{2018Nisini}). In this Section, we explore possible correlations between the observed free-free emission and the line properties, both for the low-velocity component (LVC) and high-velocity component (HVC), following the typical spectral decomposition in the [O I] line. 

\begin{figure} 
   \centering
   \includegraphics[width=0.4\textwidth]{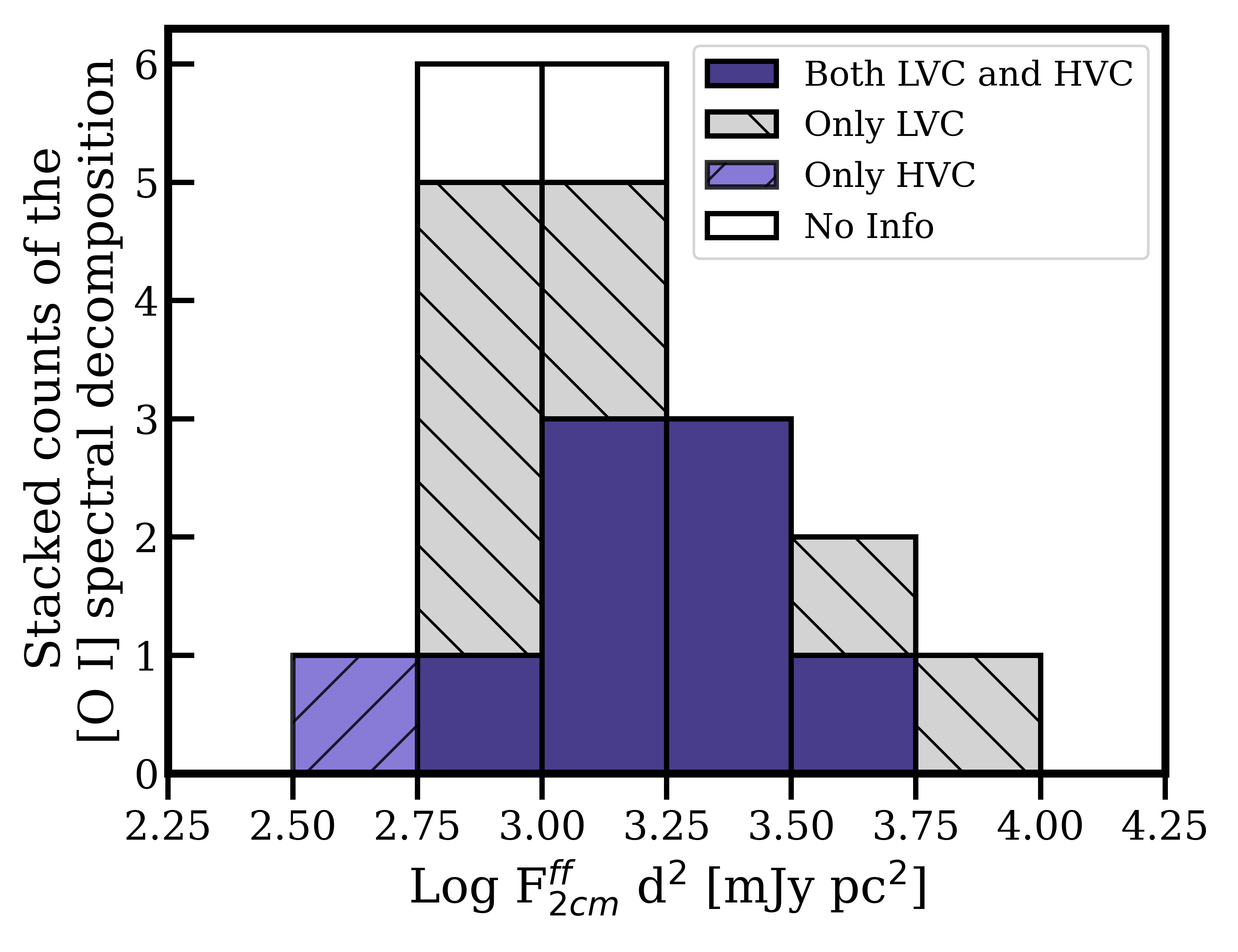}
      \caption{Histogram showing the (stacked) distribution of [O I ]6300$\r{A}$ HVC and/or LVC detected in the sample.}
    \label{fig:OI_counts}  
\end{figure}

\begin{figure} 
\centering
   \includegraphics[width=0.4\textwidth]{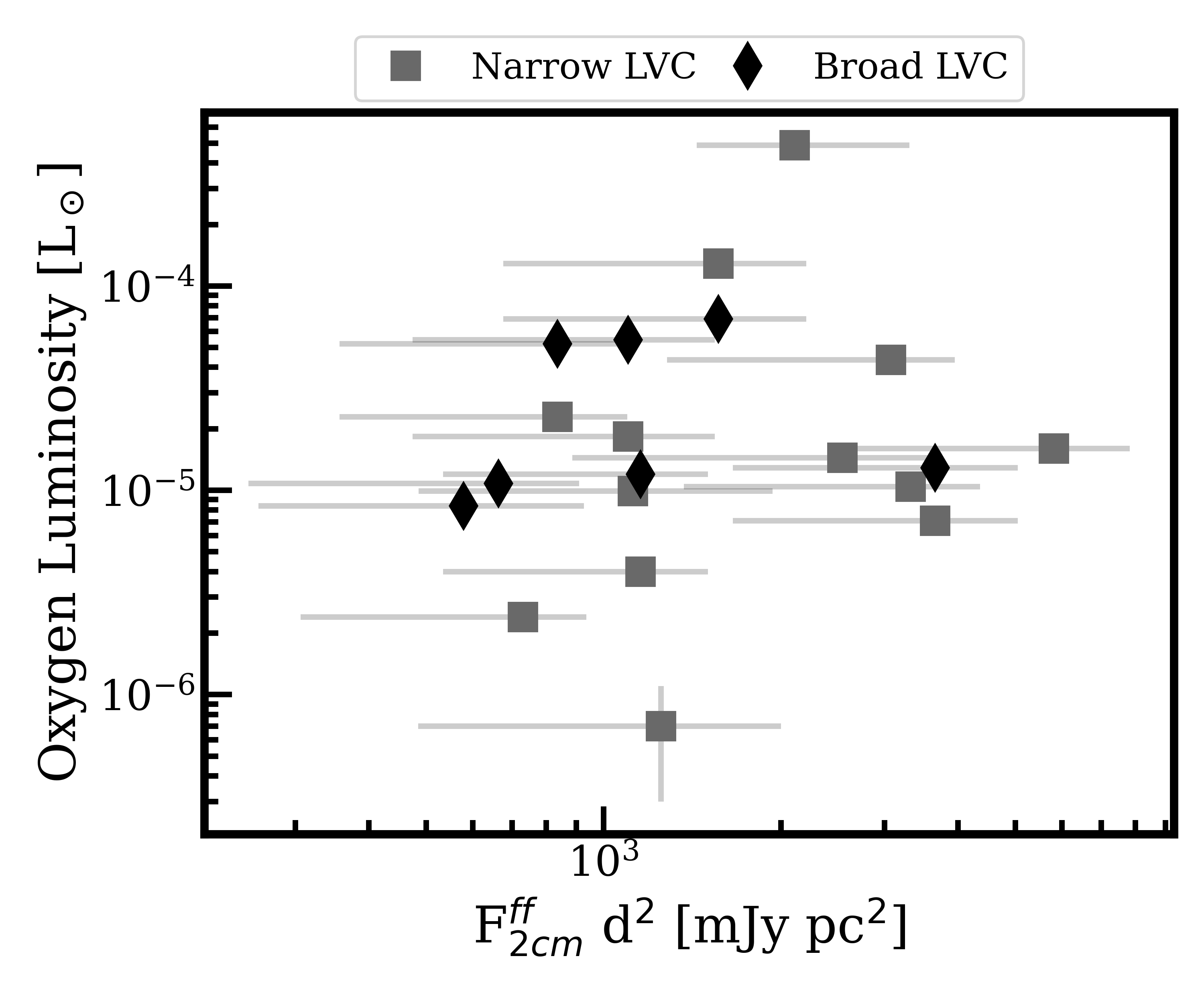}
   \centering
   \includegraphics[width=0.4\textwidth]{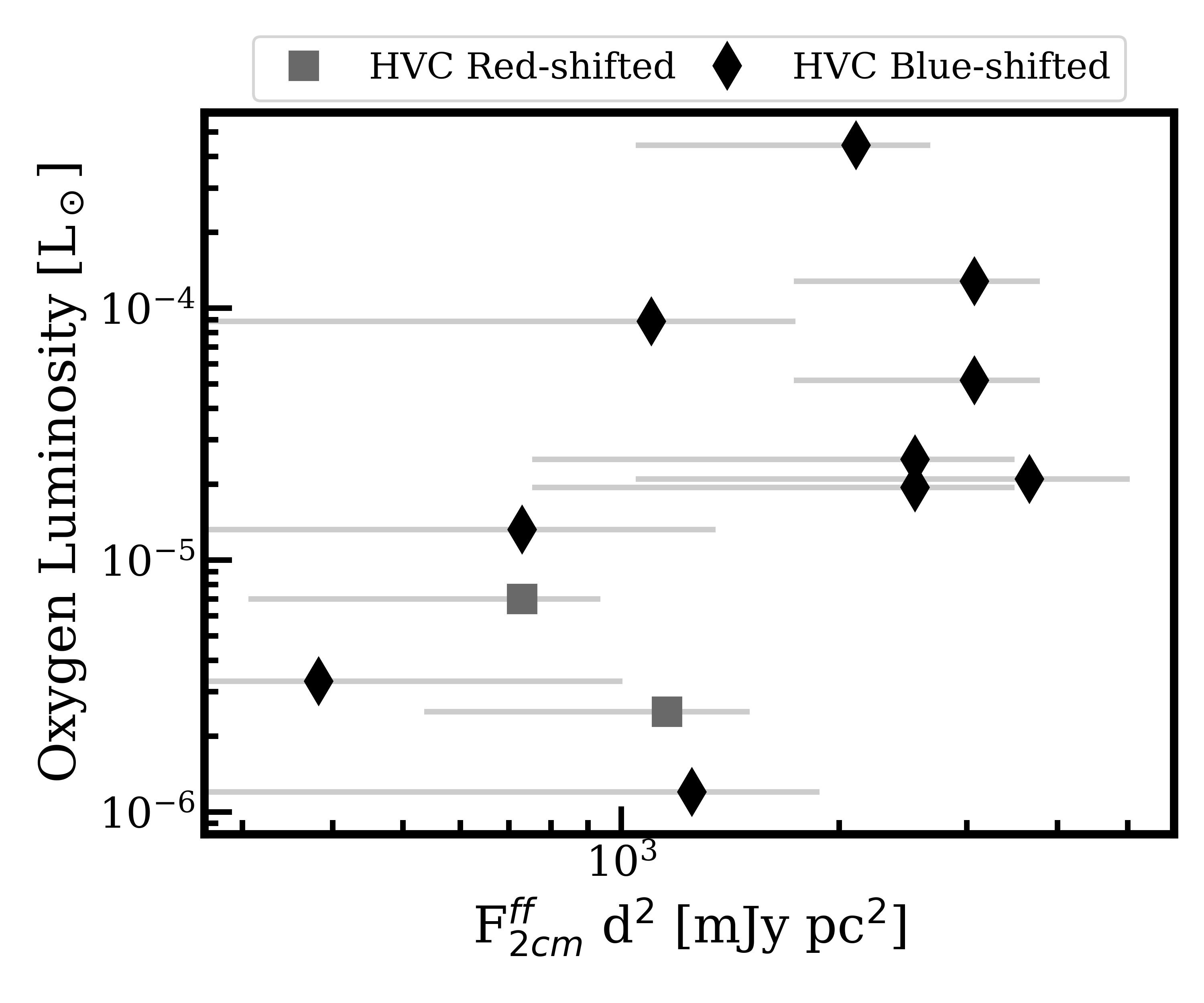}
    \caption{[O I]6300 $\r{A}$ luminosity for the HVC (\textit{bottom}) and LVC (\textit{top}) as a function of the free-free luminosity. Pearson coefficient in both plots is $|r|<0.1$.}
    \label{fig:OI_luminosity}  
\end{figure}

As reported in \cite{2024Nisini} and \cite{2016Simon}, the forbidden line [O I]6300$\r{A}$ is detected in 17 out of 19 targets. As shown in Figure \ref{fig:OI_counts}, both HVCs and LVCs are detected in 8 out of 17 targets; LVC-only is detected in 8 out of 17 targets;  HVC-only in one target. No information on the [O I] for CIDA 9 and HO Tau has been found in the literature. We explored possible correlations between the free-free luminosity and the [O I]6300$\r{A}$ line properties, specifically the line peak velocity, the line FWHM, the total luminosity, and the mass-loss rate in the jet inferred from the HVC properties. No correlation is found, neither for the LVCs nor the HVCs.
As an example, in Figure \ref{fig:OI_luminosity}, we show the oxygen luminosity as a function of the free-free luminosity (Pearson coefficient $|r|<0.1$ in all cases). Implications of the absence of these correlations are discussed in Section \ref{discussion}.

\subsubsection{Ionized mass loss rate of the jet emission}

Assuming that free-free emission is associated with gas from an ionized jet and following the geometrical model of the free-free emission from a jet by \cite{1986Reynolds}, we estimated the ionized mass loss rate $\dot{M}_\text{i}$ associated with the 2 cm free-free flux, using (see also \cite{2024Rota} and references therein):

\begin{equation}\label{eq:Mion}
\begin{split}
    \left(\frac{\dot M_\text{i}}{10^{-6} M_\odot \text{yr}^{-1}} \right) & = 0.108 \left[   \frac{(2-\alpha)(0.1+\alpha)}{1.3-\alpha} \right]^{0.75} \\
  & \times \left[ \left(\frac{F_\nu}{\text{mJy}}\right)\left(\frac{\nu}{10 \text{GHz}}\right)^{-\alpha} \right]^{0.75} \left( \frac{v_\text{jet}}{200 \text{km s}^{-1}} \right)  \\
  & \times  \left( \frac{\nu_\text{m}}{10 \text{GHz}} \right)^{0.75\alpha - 0.45} \left( \frac{\theta_0}{\text{rad}}\right)^{0.75} (\sin i)^{-0.25} \\
  & \times \left( \frac{d}{\text{kpc}} \right)^{1.5} \left( \frac{T}{10^4 \text{K}}\right)^{-0.075},
\end{split}
\end{equation}
where $\alpha$ is the spectral index, $F_\nu$ is the continuum flux at the frequency $\nu$ (2 cm in our case), and $d$ is the distance to the source. The injection opening angle of the jet $\theta_0$ is approximated as $2 \arctan(\theta_\text{min}/\theta_\text{maj})$ with the ratio between the minor and major axis of the jet $\theta_\text{min}/\theta_\text{maj}$ assumed equal to 0.5 (see \citealp{2021Kavak}). A temperature of $T=10^4$ K is adopted for the ionized gas and the jet is assumed to be fully ionized.
Since the value of the turnover frequency has not been determined directly from observations yet (\citealp{2018Anglada}) and there is not a significant dependence of the ionized mass loss rate on this value, we assumed a turnover frequency of $\nu=40$ GHz. This assumption is supported by the analysis of \cite{2025Garufi}, which found no curvature in the free-free emission spectrum below 40 GHz.
We assumed that the jet is perpendicular to the plane of the disk, and thus the inclination of the disk is taken as the inclination of the jet $i$ (see Table \ref{tab:sample}). Finally, we assumed that the velocity of the jet $v_\text{jet}$ is 
\begin{equation}
    \left( \frac{v_\text{jet}}{\text{km s}^{-1}} \right) \simeq 140 \left( \frac{M_*}{0.5 M_\odot} \right)^{1/2}
\end{equation}
(see the discussion in \citealp{2018Anglada}).

\begin{figure} 
   \centering
   \includegraphics[width=0.4\textwidth]{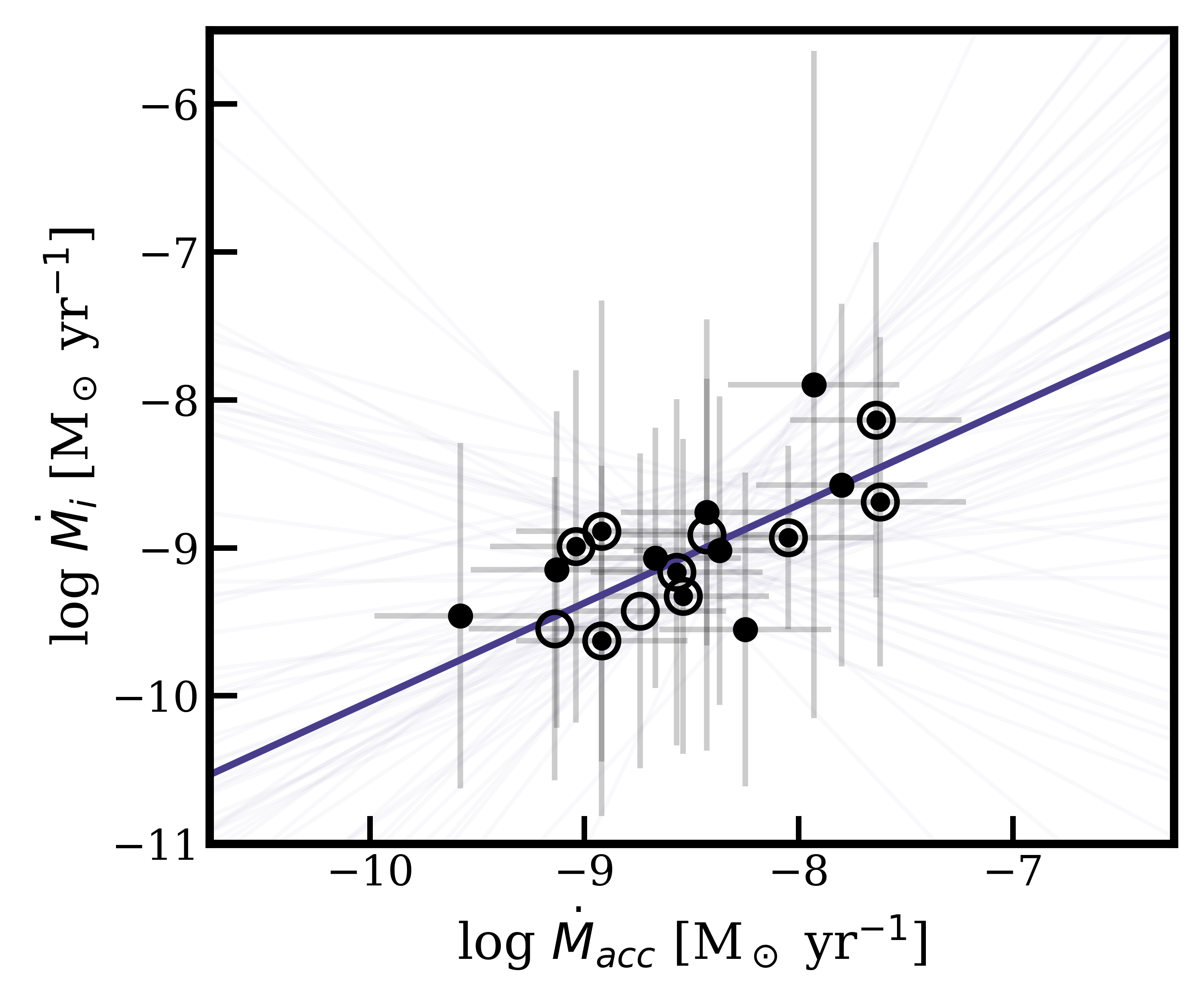}
      \caption{Ionized mass loss rate inferred from the free-free emission flux at 2 cm as a function of accretion rate onto the central star. Compact disks ($\cdot$), sub-structured disks ($\odot$), and transition disks ($\circ$) are shown.}
    \label{fig:Mloss_garufi}  
\end{figure}

Figure \ref{fig:Mloss_garufi} shows the ionized mass loss rate, as inferred from the 2 cm free-free emission, as a function of the accretion rate onto the central star, estimated for all the disks in the sample. Table \ref{tab:free-free} reports the estimated $\dot M_\text{i}$ values.
A strong correlation between the accretion rate and the ionized mass rate is found ($r=0.70 \pm 0.18$), with no particular sign of segregation between the compact and structured disks and with a median of $\dot M_{i}/\dot M_\text{acc} = 0.29$. As we will discuss in the following section, this positive correlation suggests that jets and/or ionized MHD winds are one of the main drivers of accretion onto the star in evolved YSOs. 

\section{Discussion}\label{discussion}

\subsection{Origin of free-free emission in YSOs with full disks}

We have explored various correlations between the observed free-free emission and other observables, to determine the origin of the free-free emission in the disk sample of \cite{2025Garufi}.
Considering the lack of correlation between the free-free emission and the X-ray (Figure \ref{fig:freefreeVSxray_pascucci}) and the strong correlation between the ionized mass-loss rate (inferred from the free-free emission) and the accretion rate (Figure \ref{fig:Mloss_garufi}), the free-free emission most likely originates from ionized gas associated with a jet or MHD-disk wind similar to the results found for transition disks by \cite{2024Rota}.
The free-free emission could not be spatially resolved in full disks, as was the case for transition disks, and, thus, it could not be constrained to the innermost disk regions. However, the large number of data points from the photometry up to 2 cm still allowed for a full disentangling of the free-free emission from the thermal dust emission at shorter wavelengths (\citealp{2025Garufi}).
The flat positive spectral indices suggest partially optically thick free-free emission, which is commonly found in collimated winds (\citealp{2018Anglada}).
Also, no significant differences between the compact and the sub-structured disks (defined full disks in this work) have been found in the correlation with the accretion rate. This implies that MHD-disk winds may play an important role in disk evolution, regardless of the dust substructures.
The absence of correlations between any oxygen line properties and the free-free emission (Figure \ref{fig:OI_luminosity}) suggests that the detected free-free emission traces a different part of the jet/MHD-wind than the more neutral oxygen line. Another possibility is that other contributions contaminate the detected free-free emission. Non-thermal emissions, such as the gyro-synchrotron emission, which cannot be subtracted by the SED fitting method applied by \cite{2025Garufi}, may contribute to a not-determinable fraction of the flux, making it impossible to observe a correlation between the [O I] properties and the `pure' free-free emission. However, this latter possibility is unlikely since these other types of emission are likely to vary from target to target, also affecting the correlation between the free-free flux and the accretion rate.

\subsection{Comparison between transition disks and full disks}\label{comparison}

To assess whether the mechanisms behind the free-free emission are universal in all classes of protoplanetary disks, we compare the samples from \cite{2025Garufi} (to which we added TW Hya and HD163296 from the literature, \citealp{2021Macias}; \citealp{2022Guidi}) and \cite{2024Rota}. In this comparison, CIDA 9, GI Tau, and IP Tau, analyzed in \cite{2025Garufi}, are classified as transition disks. The question is whether the mechanisms involved are different for transition disks than for full disks, considering the presence of the large inner cavity and the possible origin of the cavity \citep{2022Martel,2023vanderMarel}.

\begin{figure} 
   \centering
   \includegraphics[width=0.4\textwidth]{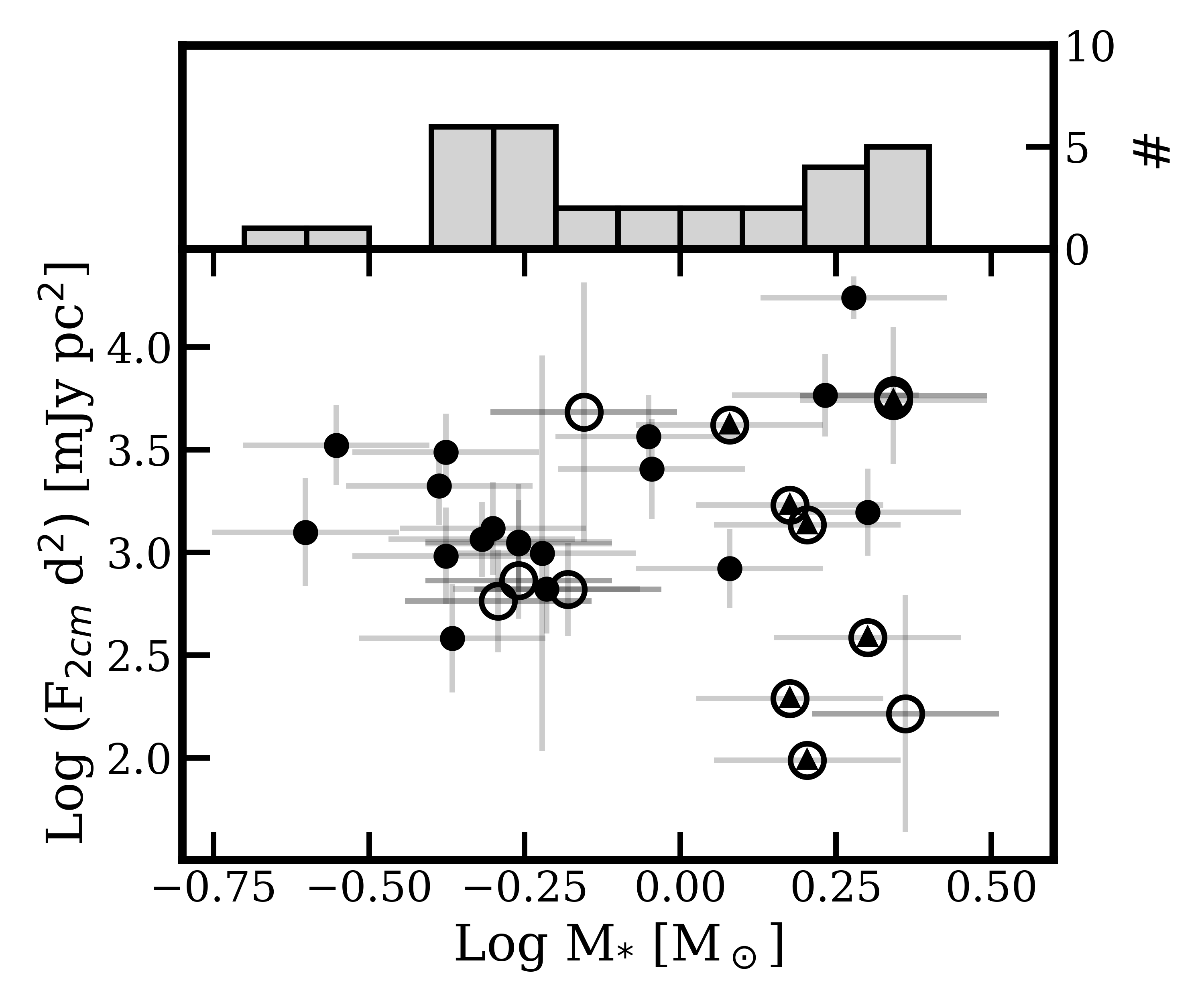}
   \centering
   \includegraphics[width=0.4\textwidth]{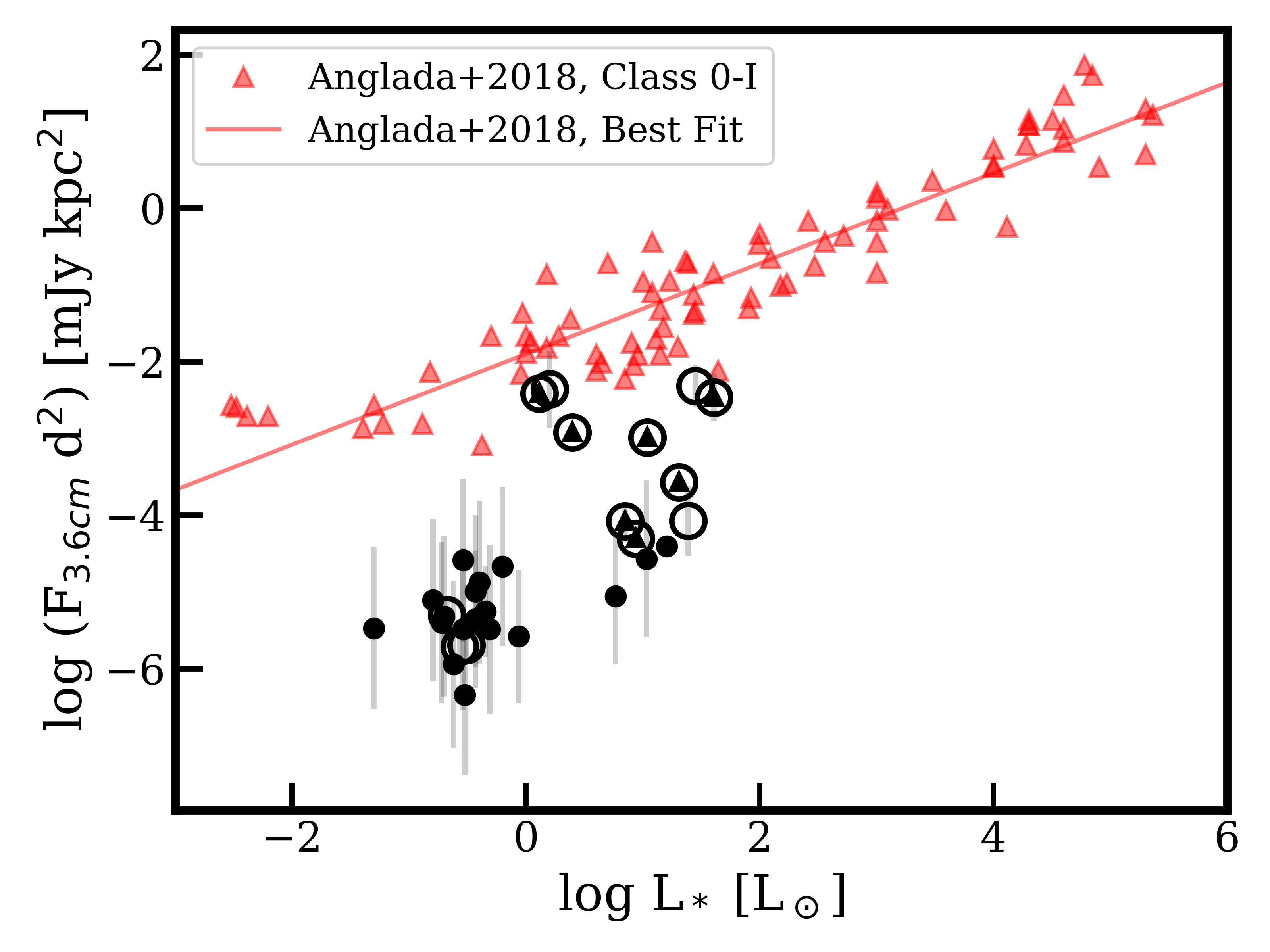}
      \caption{\textit{Top:} Stellar mass as a function of the free-free luminosity at 2 cm estimated by \cite{2025Garufi} and by \cite{2024Rota}. The plot also reports the histogram counts of the stellar mass. \textit{Bottom:} 3.6 cm luminosity as a function of the stellar/bolometric luminosity for the two analyzed sample and for Class 0-I radio jets (\citealp{2018Anglada}). In both panels, full disks ($\cdot$) and transition disks ($\circ$) are shown. The luminosities of most of transition disks are lower limits (triangles) to the real values since the spectral indices used to extrapolate them are upper limits (see text for further details).}
    \label{fig:Mstar_both}  
\end{figure}

In Figure \ref{fig:Mstar_both}, we show the free-free emission as a function of stellar properties for both samples.
For the transition disks analyzed in \cite{2024Rota}, a complete study of the dust contribution at low frequency was possible only in three disks (GG Tau, HD 100546, and HD 142527). In all the other cases, the estimated spectral indices of the free-free emission are upper limits to the real values (Table \ref{tab:free-free} and \cite{2024Rota} for further details). The extrapolated free-free luminosities at 2 cm and 3.6 cm for these disks are thus lower limits.
As shown in the top panel, the sample of transition disks is biased towards highly massive and highly accreting targets, with accretion rates spanning from $10^{-7.5}$ to $10^{-6} \text{M}_\odot\text{/yr}$. All disks in the two samples are associated with Class II sources and are found to be under-luminous when compared to younger radio jets detected in Class 0-I YSOs (\citealp{2018Anglada}, bottom panel in Figure \ref{fig:Mstar_both}). 
This is expected since disks around Class II sources are more evolved than those associated with Class 0-I YSOs. Their total luminosity is more affected by the stellar contribution than that of younger sources, whose accretion component of the luminosity is correlated with the flux of the radio jet (\citealp{2018Anglada}). Therefore, the luminosity of Class II radio jets is expected to correlate with only a fraction of the bolometric luminosity (\citealp{2024Palau, 2023Betti, 2021RiazBally}). Moreover, as shown by the figure, since the sample of transition disks is biased towards higher accretors, it is expected to be biased toward more massive disks and thus more luminous compared to less massive disks around Class II YSOs.

\begin{figure} 
   \centering
   \includegraphics[width=0.4\textwidth]{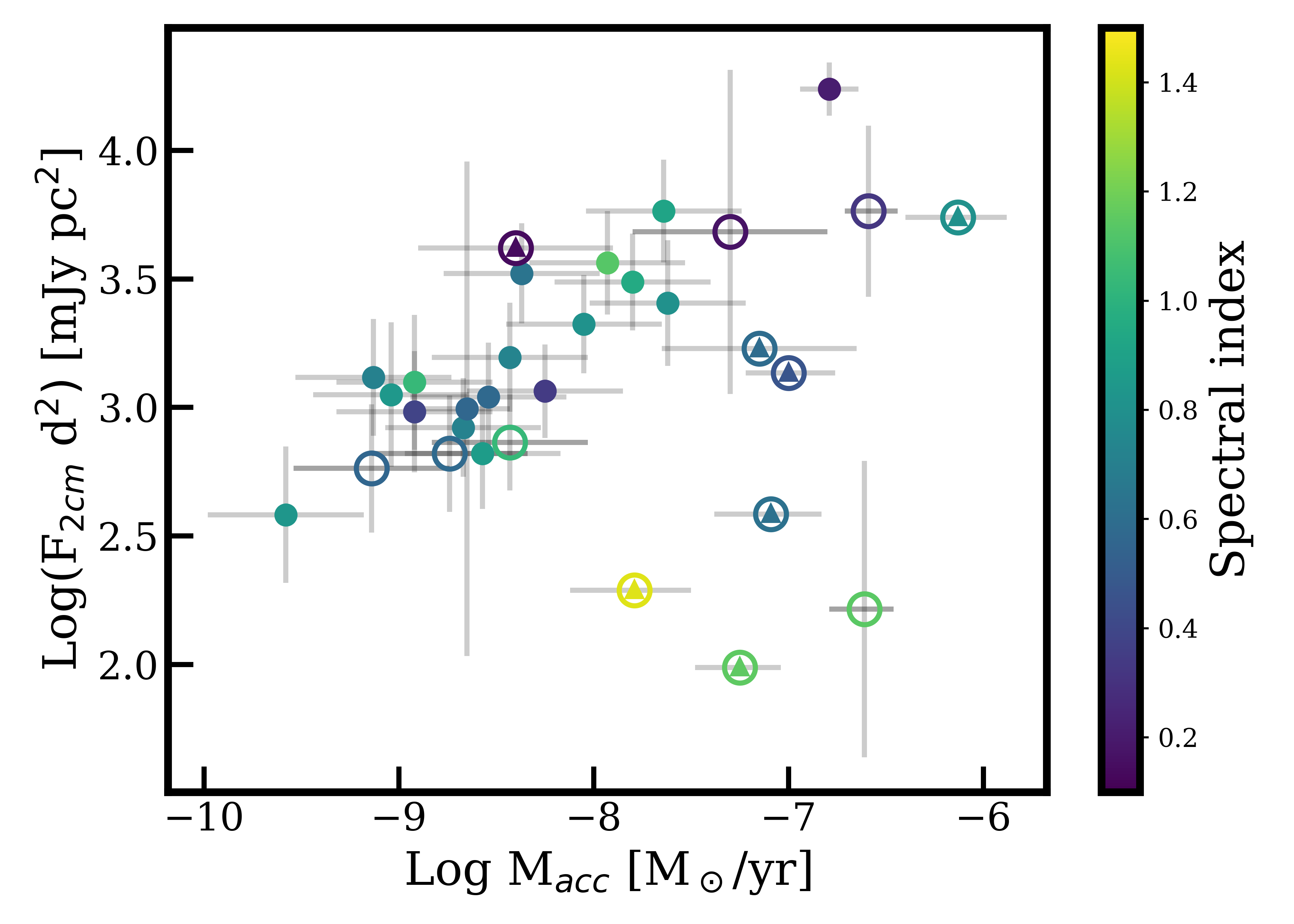}
   \includegraphics[width=0.4\textwidth]{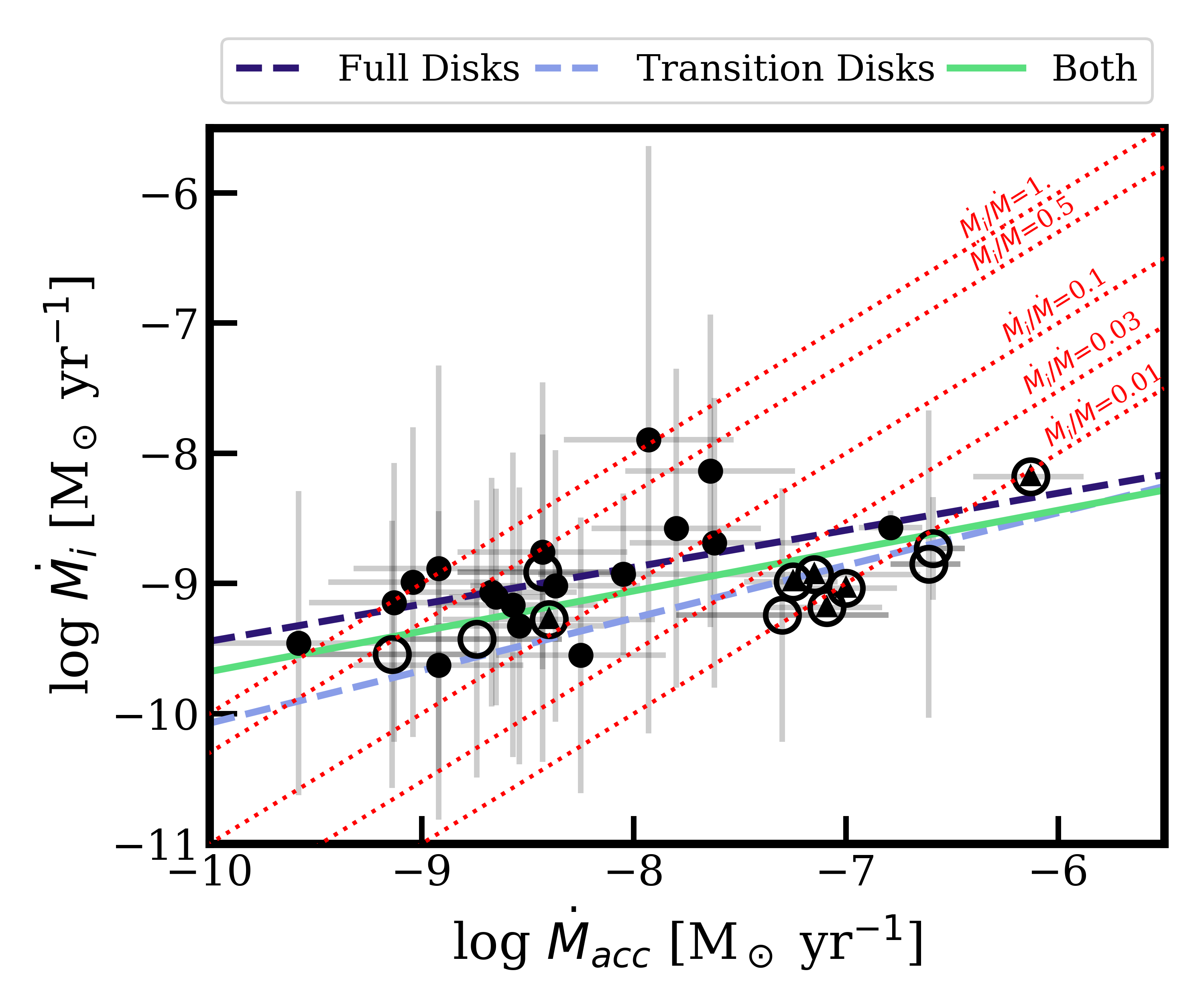}
      \caption{Free-free luminosity at 2 cm (\textit{top}) and ionized mass loss rate (\textit{bottom}) as a function of accretion rate for full disks ($\cdot$) and transition disks ($\circ$). The luminosities at 2 cm of most of transition disks are lower limits (triangles) to the real values since the spectral indices used to extrapolate them are upper limits (see text for further details). The colorbar shows the spectral index of the emission. In the bottom panel, HD100453 is missing since it was not possible to calculate the ionized mass loss rate (see Table \ref{tab:free-free} and \citealp{2024Rota}).}
    \label{fig:freefreeVSacc_both}  
\end{figure}

The top panel in Figure \ref{fig:freefreeVSacc_both} shows the free-free luminosity at 2 cm as a function of the stellar accretion rate for both transition and full disks. 
%For the transition disks analyzed in \cite{2024Rota}, a complete study of the dust contribution at low frequency was possible only in three cases (GG Tau, HD 100546, and HD 142527). In all the other cases, the estimated spectral indices of the free-free emission are upper limits to the real values (Table \ref{tab:free-free} and \cite{2024Rota} for further details). 
%\textbf{As mentioned above,} the extrapolated free-free luminosities at 2 cm for \textbf{most of transition} disks are lower limits. 
%Therefore, since this caveat affects most of the disks in the transition disk sample, 
Since the extrapolated free-free luminosities at 2 cm for most of transition disks are lower limits, a quantitative comparison between the transition and full disks is challenging. However, considering the lower limits, the two samples hint at a similar trend of the 2 cm free-free luminosity with the stellar accretion rate. More observations of transition disks at low frequencies are required to disentangle the dust emission close to the star from the free-free emission, allowing one to investigate possible differences between the transition and full disk samples (see Section \ref{implications} for more details).
%\aar{The two samples show a similar trend of the 2 cm free-free luminosity with the stellar accretion rate, with the transition disks showing a shallower correlation than the full disks. The interpretation of this shallower correlation is affected by the biases in our sample. Since the sample of transition disks is biased toward high accretors and the sample of full disks is missing the higher tail of the accretion rate distribution, it is not possible to assess whether this shallower correlation applies only to stronger accretors or to transition disks in general (see Section \ref{implications} for more details).}

Assuming that the free-free emission is associated with gas from an ionized jet and using Equation (2), we compare the ionized mass loss rate $\dot{M}_\text{i}$ associated with the 2 cm free-free flux in the two samples.
As shown by the bottom panel of Figure \ref{fig:freefreeVSacc_both} and as reported in Table \ref{tab:free-free}, $\dot{M}_\text{i}$ spans from $10^{-10}$ to $10^{-8} \text{M}_\odot\text{/yr}$ and shows a similar trend with the accretion rate in both samples. A strong correlation between the accretion rate and the ionized mass rate is found both for the transition disks (blue dashed line, see also \citealp{2024Rota}) and the full disks (purple dashed line, see also Figure \ref{fig:Mloss_garufi}). Moreover, a hint of correlation is found when combining the two samples (green solid line). The found relationships are summarized in the following\footnote{The linear regression was performed with  \texttt{linmix} (\citealp{2007Kelly})}: 

\begin{itemize}
    \item[-] \text{transition disks only ($r=0.79\pm0.19, p=0.002$):}
$$   \left( \frac{\dot M_{i}}{M_\odot \text{\text{yr}}^{-1}} \right)_{\text{TD}} = 10^{-5.97 \pm 2.00} \left(\frac{\dot M_\text{acc}}{\text{M}_\odot \text{yr}^{-1}} \right)^{0.41\pm 0.28} \quad \text{(a)};
$$

\item[-] \text{full disks only ($r=0.64\pm0.19, p=0.004$):}
$$
   \left( \frac{\dot M_{i}}{M_\odot \text{\text{yr}}^{-1}} \right)_{\text{full}} = 10^{-6.53 \pm 2.45} \left(\frac{\dot M_\text{acc}}{\text{M}_\odot \text{yr}^{-1}} \right)^{0.29\pm 0.31}  \quad \text{(b)};
$$

\item[-] \text{both samples combined ($r=0.53\pm0.16, p=0.002$):}
$$
    \left( \frac{\dot M_{i}}{M_\odot \text{\text{yr}}^{-1}} \right)_{\text{both}} = 10^{-6.62 \pm 1.14} \left(\frac{\dot M_\text{acc}}{\text{M}_\odot \text{yr}^{-1}} \right)^{0.31\pm 0.15}  \quad \text{(c)}.
$$
\end{itemize}

The correlation reported for the transition disk sample (Equation (a)) is affected by the fact that the ionized mass loss rate in most of these targets is inferred from lower limits on the 2 cm free-free flux and its exact shape should be analyzed together with a complete study of the dust emission in the cavity of these disks. However, excluding these targets from the combined sample of full and transition disks leads to a consistent correlation with the one reported in Equation (c). Both the transition disks, the full disks, and the combined sample show fully consistent slopes within the errorbars.

These correlations suggest that the detected free-free emission is in both samples associated with a similar mechanism, i.e., ionized gas close to the star from an MHD-wind and/or jet, and reveals a strong relation between the accretion and outflow properties. 
One possibility is that, in both transition and full disks, jets and/or MHD-winds are one of the main drivers of accretion onto the central star. On the other hand, other processes, such as turbulence (e.g., \citealp{2023ManaraPPVII}), might be the main drivers of the accretion, which causes a shock that consequently launches the outflow.

\subsection{Implications for angular momentum transport in disks}\label{implications}

\begin{figure} 
   \centering
   \includegraphics[width=0.5\textwidth]{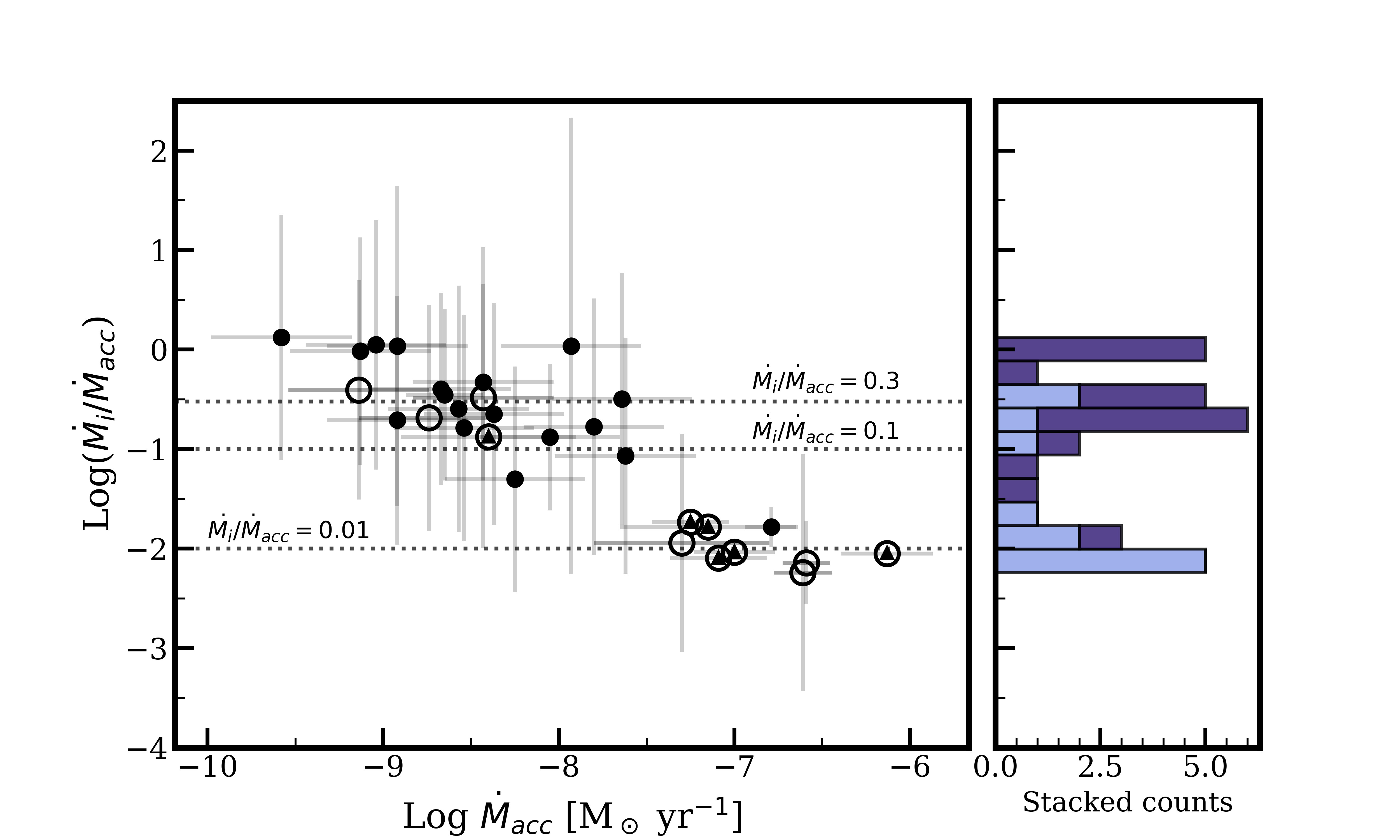}
    \centering
   \includegraphics[width=0.4\textwidth]{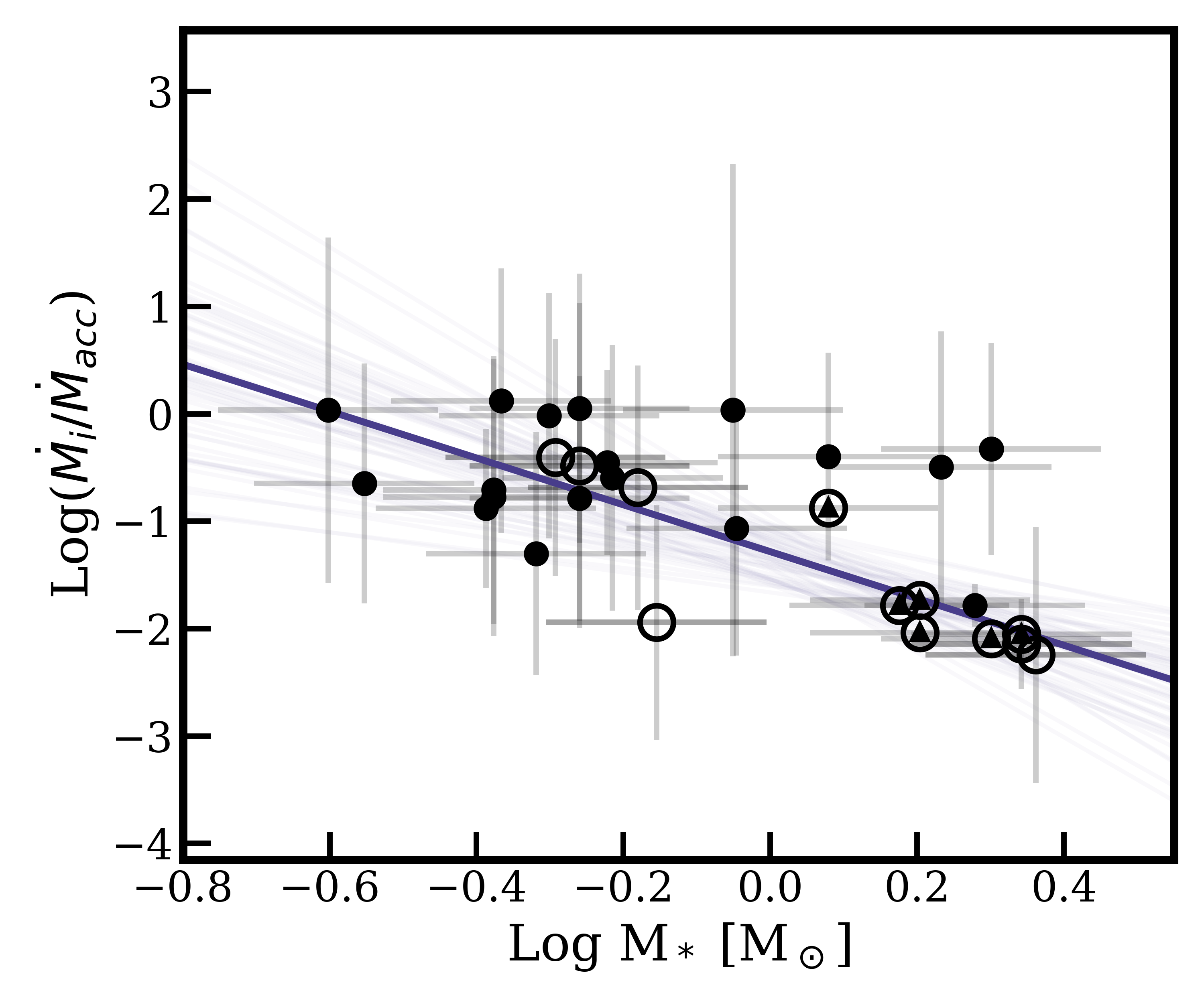}
      \caption{\textit{Top:} Jet efficiency at 2 cm as a function of the accretion rate. The plot also reports the histogram counts of the $\dot M_{i}/\dot M_\text{acc}$ ratio for transition (blue) and full disks (purple). \textit{Bottom:} Jet efficiency at 2 cm as a function of the stellar mass. }
    \label{fig:efficiecy}  
\end{figure}

The ratio between the ionized mass loss rate and the accretion rate $\dot M_{i}/\dot M_\text{acc}$ can be used to characterize the efficiency $\xi$ of the outflow launch. 
If all the MHD disk wind is ejected from a small region at the footpoint radius, then  $M_{i}/\dot M_\text{acc} \sim \lambda^{-1} $, where $\lambda$ is the magnetic lever arm (e.g., \citealp{1992Pelletier&Pudritz}; \citealp{2023PascucciPPVII}). A more heavily mass-loaded MHD-disk wind (large $\xi$) will necessarily have a smaller $\lambda$, i.e. a smaller angular momentum per unit mass, and viceversa (\citealp{2023PascucciPPVII}). Inside the MRI-active region ($0.05-0.5$ au), $\lambda \sim 5$ is expected from global simulations (\citealp{2021Jacquemin-Ide}). 
As shown by the top panel in Figure \ref{fig:efficiecy}, $\dot M_{i}/\dot M_\text{acc}$ spans from $0.01$ to $1.3$ in the combined sample of transition and full disks (median $0.18$), with the transition disks showing a median $\dot M_{i}/\dot M_\text{acc}$ of $0.014$ and the full disks a median of $0.29$. 
Assuming a jet origin, this range of values is consistent with $1/\lambda=0.2$ expected from global simulations of jet-launching in the MRI-active region (\citealp{2021Jacquemin-Ide}).
Stronger accretors show a tendency toward lower jet efficiencies, with values of  $\dot M_{i}/\dot M_\text{acc}<0.1$, while sources with low accretion rates (i.e., $\dot M_\text{acc} < 10^{-8} \text{M}_\odot\text{/yr}$) show a ratio larger than 0.1. These results are consistent with findings from the HVC of the [O I]6300 $\r{A}$ line in a sample of 131 disks in Lupus, Chamaeleon, and $\sigma$Orionis star-forming regions (\citealp{2018Nisini}). 

Moreover, the $\dot M_{i}/\dot M_\text{acc}$ ratios in the transition and full disk samples are unlikely drawn by the same distribution, as confirmed by the Kolmogorov-Smirnov (K-S) two-sided test for the null hypothesis that the two samples are drawn from the same continuous distribution (p-value of $\sim 0.0015$). However, this result is probably due to the bias towards highly accreting disks which affects the transition disk sample and is probably affected by the fact that most of the extrapolated free-free fluxes at 2 cm for transition disks are lower limits (see Table \ref{tab:free-free}). The tendency of weaker accretors toward larger values of $\dot M_{i}/\dot M_\text{acc}$ suggests that the jet efficiency in stronger accretors may show a shallower correlation with the accretion rate. Another possibility is that transition disks in general show a different efficiency in transforming accretion into outflow than full disks, probably in turn related to the efficiency of transport of material through the cavity.
%As for the correlation between the free-free luminosity and the accretion rate (Figure \ref{fig:freefreeVSacc_both}), the jet efficiency in stronger accretors may show a shallower correlation with the accretion rate. Another possibility is that transition disks in general show a different efficiency in transforming accretion in outflow than full disks, probably in turn related to the efficiency of transport of material through the cavity.
New observations of transition disks covering the lower hand of the accretion rate distribution and of full disks in higher accreting targets are needed to assess how the jet efficiency in the two samples differs.

Finally, the $\dot M_{i}/\dot M_\text{acc}$ ratio depends on the stellar mass, as shown by the bottom panel in Figure \ref{fig:efficiecy}. An anti-correlation between these two quantities is found for the combined sample of full and transition disks, with a Pearson coefficient $r=-0.65\pm0.14$. The correlation holds for the sample of transition disks alone ($r=-0.84\pm0.17$), while it disappears when considering the sample of full disks alone ($r=-0.24\pm0.24$). This strong anti-correlation, driven by the sources with high accretion rates, might suggest that the found correlation between the ionized mass loss rate and the accretion rate might be driven by the stellar mass. As shown by Equations (2) and (3), $\dot M_{i}$ depends on the stellar mass through the jet velocity $v_\text{jet}$. To test if the correlation between $\dot M_{i}$ and $\dot M_\text{acc}$ is driven by the stellar mass, we fixed the jet velocity in Equation (2) to $200 \text{km/s}$ and recalculated the ionized mass loss rate. Even with this assumption, the correlation between the ionized mass loss rate and the accretion rate holds both for full and transition disks, showing that it is unlikely driven by the stellar mass dependency.  
Therefore, the anti-correlation between the jet efficiency and the stellar mass is likely driven by the steep correlation between the stellar accretion rate and the stellar mass (see Section \ref{results}).

%--------------------------------------------------------------------

\section{Conclusions}\label{conclusions}

In this work, we analyzed a sample of 31 YSOs for which multi-wavelength observations at millimeter and centimeter wavelengths are available in the literature.
We first analyzed the free-free emission recently detected in a sample of Class II YSOs with disks in Taurus, reported by \cite{2025Garufi}. 
No correlation between the detected free-free emission and either the X-ray and the [O I]6300 \r{A} line properties is found. On the other hand, we found a strong correlation between the detected free-free luminosity and the accretion luminosity and between the ionized mass loss rate (as inferred from the free-free emission) and the stellar accretion rate. We thus conclude that the free-free emission in those YSOs most likely originates from ionized gas associated with a jet or MHD-disk wind and traces a different part of the jet/MHD-wind than the more neutral oxygen line.

We then compared our findings on full disks with the transition disk sample from \cite{2024Rota}. Our results can be summarized as follows:

\begin{itemize}
    \item Considering the lower limits, the free-free luminosity detected in transition disks suggests a similar trend with the accretion rate as observed in full disks.%a shallower correlation with the accretion rate than in full disks.
    \item A similar correlation between the ionized mass loss rate (inferred from the free-free emission) and the accretion rate is found for both samples.
    
    \item Sources with high accretion rates ($>10^{-8}\text{M}_\odot\text{/yr}$) show a tendency toward lower jet efficiency, with $\dot M_{i}/\dot M_\text{acc} < 0.1$, while weaker accretors shows ratios larger than 0.1.
    \item A strong correlation driven by the high accretors is found between $\dot M_{i}/\dot M_\text{acc}$ and the stellar mass.
\end{itemize}

The detected free-free emission is in both samples likely associated with a similar mechanism, i.e., ionized gas close to the star from an MHD-wind and/or jet. %The shallower correlations found for the sample of transition disks 
The lower median values of $\dot M_{i}/\dot M_\text{acc}$ found for the transition disk sample hint at a different efficiency in transforming accretion into outflow than full disks, probably related to the transport of material through the cavity. However, since a complete study of the dust contribution at low frequency was not possible for all the transition disks, more long-wavelength observations of transition disks are required to quantitatively investigate the free-free emission contribution in those targets.
Moreover, since the sample of transition disks is biased toward high accretors, new observations of highly accreting full disks and less accreting transition disks, are needed to assess whether the hinted shallower correlation between the jet efficiency and the accretion rate holds in general for transition disks or whether it applies to strong accretors only. This will help in understanding how efficiently the material is transported through the large cavities observed in transition disks.

\begin{acknowledgements}
%This paper makes use of the following ALMA data: ALMA is a partnership of ESO (representing its member states), NSF (USA) and NINS (Japan), together with NRC (Canada), MOST and ASIAA (Taiwan), and KASI (Republic of Korea), in cooperation with the Republic of Chile. The Joint ALMA Observatory is operated by ESO, AUI/NRAO and NAOJ. The PI acknowledges assistance from Allegro, the European ALMA Regional Center node in the Netherlands.  
The authors thank the referee for their useful comments that
have contributed to improving the manuscript. CC-G acknowledges support from UNAM DGAPA PAPIIT grant IG101224 and from CONAHCyT Ciencia de Frontera project ID 86372. SF is funded by the European Union (ERC, UNVEIL, 101076613), and acknowledges financial contribution from PRIN-MUR 2022YP5ACE. Views and opinions expressed, however, are those of the author(s) only and do not necessarily reflect those of the European Union or the ERC. Neither the European Union nor the granting authority can be held responsible for them.

\end{acknowledgements}

\bibliography{bibliography.bib}

\end{document}